\documentclass[final]{siamltex}
\usepackage{amsmath, amssymb, amsfonts}
\usepackage{graphicx,color,caption,subcaption}
\usepackage{diagbox}
\usepackage[ruled]{algorithm2e}
\usepackage{algorithmic}
\usepackage{enumitem}
\usepackage{multirow}
\usepackage{showlabels}

\newcommand{\bvec}[1]{\mathbf{#1}}

\newcommand{\tabincell}[2]{\begin{tabular}{@{}#1@{}}#2\end{tabular}}

\newcommand{\mc}[1]{\mathcal{#1}}
\newcommand{\mf}[1]{\mathfrak{#1}}
\newcommand{\mcV}{\mathcal{V}}

\newcommand{\xc}{\mathrm{xc}}
\newcommand{\vr}{\bvec{r}}
\newcommand{\vF}{\bvec{F}}

\newcommand{\vR}{\bvec{R}}

\newcommand{\ud}{\,\mathrm{d}}

\newcommand{\KS}{\mathrm{KS}}

\newcommand{\abs}[1]{\lvert#1\rvert}

\newcommand{\wt}[1]{\widetilde{#1}}
\newcommand{\fhxc}{\mathfrak{f}_{\mathrm{hxc}}}

\newcommand{\Or}{\mathcal{O}}

\newcommand{\I}{\imath}

\newcommand{\REV}[1]{{#1}}

\title{Split representation of adaptively compressed polarizability operator}

\author{
Dong An\thanks{Department of Mathematics, University of California, Berkeley, Berkeley, CA 94720. Email: \texttt{dong\_an@berkeley.edu}}
\and
        Lin Lin\thanks{Department of Mathematics, University of California, Berkeley, Berkeley, CA 94720 and Computational Research Division, Lawrence Berkeley National Laboratory, Berkeley, CA 94720. Email: \texttt{linlin@math.berkeley.edu}}
\and Ze Xu\thanks{Department of Mathematics, University of California, Berkeley, Berkeley, CA 94720. Email: \texttt{zexu@math.berkeley.edu}}
}

\begin{document}

\maketitle

\begin{abstract}
The polarizability operator plays a central role in density functional
perturbation theory and other perturbative treatment of first principle
electronic structure theories.  The cost of computing the polarizability
operator generally scales as $\Or(N_{e}^4)$ where $N_e$ is the number of
electrons in the system. The recently developed adaptively compressed
polarizability operator (ACP) formulation [L. Lin, Z. Xu and L. Ying,
Multiscale Model. Simul. 2017] reduces such complexity to
$\Or(N_{e}^3)$ in the context of phonon calculations with a large basis
set for the first time, and demonstrates its effectiveness for model
problems. In this paper, we improve the performance of the ACP
formulation by splitting the polarizability into a near singular
component that is statically compressed, and a smooth component that is
adaptively compressed. The new split representation maintains the
$\Or(N_e^3)$ complexity, and accelerates nearly all components of
the ACP formulation, including Chebyshev interpolation of energy levels,
iterative solution of Sternheimer equations, and convergence of the Dyson equations. For simulation of real materials, we discuss how to incorporate nonlocal pseudopotentials and finite temperature effects.
We demonstrate the effectiveness of our method using one-dimensional model problem in insulating and metallic regimes, as
well as its accuracy for real molecules and solids.
\end{abstract}



\pagestyle{myheadings}
\thispagestyle{plain}
%\markboth{L. Lin, Z. Xu and L. Ying}{Adaptively Compressed Polarizability Operator for Large Scale Phonon Calculations}

%%%%%%%%%%%%%%%%%%%%%%%%%%%%%%%%%%%%%%%%%%%%%%%%%%%%%%%%%%%%%%%%%%%%%%%%%%%%%%%%
\section{Introduction}\label{sec:intro}

Density functional perturbation theory
(DFPT)~\cite{BaroniGiannozziTesta1987,GonzeLee1997,BaroniGironcoliDalEtAl2001,CancesMourad2014}
studies the response of a quantum system under small perturbation, where
the quantum system is described at the level of first principle
electronic structure theories such as Kohn-Sham density functional
theory (KSDFT)~\cite{HohenbergKohn1964,KohnSham1965}.  One important
application of DFPT is the calculation of vibration properties such as
phonons, which can be further used to calculate many physical properties
such as infrared spectroscopy, elastic neutron scattering, specific
heat, heat conduction, and electron-phonon interaction related behaviors
such as superconductivity (see \cite{BaroniGironcoliDalEtAl2001} for a
review). DFPT describes vibration properties through a polarizability
operator, which characterizes the linear response of the electron
density with respect to the perturbation of the external potential. More
specifically, in vibration calculations, the polarizability operator
needs to be applied to $d\times N_A\sim \Or(N_e)$ perturbation vectors,
where $d$ is the spatial dimension (usually $d=3$), $N_A$ is the
number of atoms, and $N_{e}$ is the number of electrons. In general the complexity for solving KSDFT is
$\Or(N_e^3)$, while the complexity for solving DFPT is $\Or(N_e^4)$. It
is possible to reduce the computational complexity of DFPT
calculations by ``linear scaling
methods''~\cite{Goedecker1999,NiklassonChallacombe2004,BowlerMiyazaki2012}. Such methods can be
successful in reducing the computational cost for systems of large sizes
with substantial band gaps, but this can be challenging for medium-sized
systems with relatively small band gaps.

\REV{The term ``phonon calculation'' usually describes the calculation
of vibration properties of condensed matter
systems. In this paper, we slightly abuse this term to refer to calculations of
vibration properties of general systems, including condensed matter
systems as well as isolated molecule clusters, since such calculations
share the same mathematical structure.}
% In the discussion below, we will slightly abuse the term ``phonon
% calculation'' to refer to calculation of vibration properties of
% condensed matter systems as well as isolated molecules.
In order to
apply the polarizability operator to $\Or(N_{e})$ vectors, we need to
solve $\Or(N_{e}^2)$ coupled Sternheimer equations. On the other hand,
when a constant number of degrees of freedom per electron is used, the
size of the Hamiltonian matrix is only $\Or(N_e)$. Hence asymptotically
there is room to obtain a set of only $\Or(N_{e})$ ``compressed
perturbation vectors'', which encodes essentially all the information of
the $\Or(N_{e}^2)$ Sternheimer equations. The recently developed
adaptively compressed polarizability operator (ACP)
formulation~\cite{LinXuYing2017} follows this route, and successfully
reduces the computational complexity of phonon calculations to
$\Or(N_{e}^3)$ for the first time. The ACP formulation does not rely on
exponential decay properties of the density matrix as in linear scaling
methods, and its accuracy depends weakly on the size of the band gap.
Hence the method can be used for phonon calculations of both insulators and semiconductors
with small gaps. 

There are three key ingredients of the ACP formulation. 1) The
Sternheimer equations are equations for shifted Hamiltonians, where each
shift corresponds to an energy level of an occupied band. Hence for a
general right hand side vector, there are $N_{e}$ possible energies
(shifts). We use a Chebyshev interpolation procedure to disentangle such
energy dependence so that there are only constant number of shifts that
is independent of $N_{e}$. 2) We disentangle the $\Or(N_{e}^2)$ right
hand side vectors in the Sternheimer equations using the recently developed interpolative separable
density fitting procedure, to compress the right-hand-side vectors. 3) We construct the polarizability operator by adaptive compression so that the operator remains low rank as well as accurate when applying to a certain set of vectors. This make it possible for fast computation of the matrix inversion using methods like Sherman-Morrison-Woodbury. In particular, the ACP  method does not
employ the ``nearsightedness'' property of electrons for insulating
systems with substantial band gaps as in linear scaling
methods~\cite{Kohn1996}. Hence the ACP method can be applied to insulators
as well as semiconductors with small band gaps.

In this paper, we introduce a generalization the ACP formulation for efficient phonon calculations of real
materials called split representation of ACP. In the split representation, the nonlocal pseudopotential is taken into account, as
well as temperature effects especially for metallic systems. The new 
split representation maintains the
$\Or(N_e^3)$ complexity, and improves all key steps in the ACP formulation, including Chebyshev interpolation of energy levels,
iterative solution of Sternheimer equations, and convergence of the Dyson equations.
 
The rest of the paper is organized as follows. Section~\ref{sec:prelim} introduces the basic formulation of KSDFT and DFPT, and reviews the formulation of ACP. Section~\ref{sec:splitACP} describes the split representation of the ACP formulation. Numerical results are presented in section~\ref{sec:numer}, followed by conclusion and discussion in section~\ref{sec:conclusion}.

\section{Preliminaries}\label{sec:prelim}

%For completeness we briefly 
%
%first provide a brief introduction to Kohn-Sham density
%functional theory (KSDFT), and density functional perturbation theory (DFPT) in the
%context of phonon calculations. To simplify our discussion, we neglect
%the spin degeneracy, temperature dependence, as well as the usage of
%nonlocal pseudopotential. We assume all orbitals $\{\psi_i(\vr)\}$ are
%real. The spatial dimension $d=3$ is assumed in the treatment of e.g.
%Coulomb interaction unless otherwise specified. We remark that such
%simplified treatment does not reduce the core difficulty of the problem.

\subsection{Kohn-Sham density functional theory}

For simplicity we consider a system of
finite size with periodic boundary conditions. This can be used to model
isolated molecular systems as well as solid state systems with the Gamma
point sampling strategy of the Brillouin zone~\cite{Martin2004}.
However, we do not explicitly take advantage of that $\{\psi_i(\vr)\}$ are
real, so that the formulation is applicable to real space and Fourier
space implementation, as commonly done in electronic structure software
packages.  The spatial dimension $d=3$ is assumed in the treatment of
e.g.  Coulomb interaction unless otherwise specified.  Since our
numerical results involve real materials and systems of both insulating
and metallic characters, we include relevant technical details such as
nonlocal pseudopotential and temperature dependence in the discussion.
Consider a system consisting of $N_A$ nuclei and $N_e$ electrons at
temperature $T=1/(k_{B}\beta)$, where $k_{B}$ is the Boltzmann
constant. In
the Born-Oppenheimer approximation, for each set of nuclear
positions $\{\vR_{I}\}_{I=1}^{N_A}$, the electrons are relaxed to their ground
state. The ground state total energy is denoted by
$E_{\mathrm{tot}}(\{\vR_{I}\}_{I=1}^{N_A})$, and can be computed
in Kohn-Sham density functional
theory~\cite{HohenbergKohn1964,KohnSham1965,Mermin1965} according to the
minimization of the following 
Kohn-Sham-Mermin energy
functional
\begin{equation}\label{eqn:KSfunc}
  \begin{split}
     &E_{\KS}(\{\psi_i\};\{\vR_{I}\}) \\
     =&\frac{1}{2} \sum_{i=1}^{\infty}
     f_{i} \int \abs{\nabla \psi_i(\vr)}^2 \ud \vr + \sum_{i=1}^{\infty}
     f_{i} \int \psi^{*}_{i}(\vr)
    V_{\mathrm{ion}}(\vr,\vr';\{\vR_{I}\}) \psi_{i}(\vr') \ud \vr \ud \vr' \\
    &
%    + \sum_{i=1}^{N_{e}} \int \psi_{i}(\vr) V_{\mathrm{nl}}(\vr,\vr';\{\vR_{I}\})
%    \psi_i(\vr') \ud \vr \ud \vr'
+\frac{1}{2} \iint v_{c}(\vr,\vr')\rho(\vr) \rho(\vr') \ud \vr \ud \vr' 
+ E_{\xc}[\rho] +  E_{\mathrm{II}}(\{\vR_{I}\})\\
& + \frac{1}{\beta} \sum_{i=1}^{\infty} \left[f_{i} \log f_{i} +
(1-f_{i}) \log (1-f_{i})\right].
  \end{split}
\end{equation}
Here the minimization is with respect to the Kohn-Sham orbitals $\{\psi_i\}_{i=1}^{\infty}$
satisfying the orthonormality condition
$\int \psi_i^*(\vr) \psi_j ( \vr) \ud \vr = \delta_{ij}$, 
as well as the occupation numbers
$\{f_{i}\}_{i=1}^{\infty}$ satisfying $0\le f_{i}\le 1$. 
%We include the $\{\vR_I\}$ dependence explicitly to facilitate the
%derivation of atomic forces. 
In Eq.~\eqref{eqn:KSfunc}, $\rho(\vr) = \sum_{i=1}^{\infty}f_{i}\left| \psi_i(\vr) \right|^2$
defines the electron density with normalization condition $\int
\rho(\vr) \ud \vr = N_{e}$.   
In the discussion below we will omit the range of indices $I,i$
unless otherwise specified. 
In Eq.~\eqref{eqn:KSfunc}, $v_c(\vr,\vr') = \frac{1}{\left| \vr -
\vr' \right|}$ defines the kernel for Coulomb interaction in
$\mathbb{R}^3$ and the corresponding term is called the Hartree energy. $V_{\mathrm{ion}}$ is a potential characterizing the
electron-ion interaction, and is independent of the electronic states
$\{ \psi_i\}$. More specifically, in a pseudopotential
approximation~\cite{Martin2004}, if we view $V_{\mathrm{ion}}$ as an
integral operator, then the kernel of $V_{\mathrm{ion}}$ can be
expressed as the summation of contribution from each atom $I$
\begin{equation}
  V_{\mathrm{ion}}(\vr,\vr';\{\vR_{I}\}) = \sum_{I} V_{\mathrm{loc},I} (\vr
  - \vR_I) \delta(\vr-\vr') + \sum_{I} V_{\mathrm{nl},I} (\vr-\vR_{I},
  \vr'-\vR_{I}).
  \label{eqn:pseudopot}
\end{equation}
Here $V_{\mathrm{loc},I}$ is called the local pseudopotential, and
$V_{\mathrm{nl},I}$ the nonlocal pseudopotential. In the
Kleinman-Bylander form~\cite{KleinmanBylander:82}, each nonlocal
pseudopotential is a low rank and symmetric operator with kernel
\begin{equation}
  V_{\mathrm{nl},I} (\vr-\vR_{I}, \vr'-\vR_{I}) = \sum_{l=1}^{L_{I}}
  \gamma_{I,l} b_{I,l}(\vr-\vR_{I}) b^{*}_{I,l}(\vr'-\vR_{I}).
  \label{eqn:nonlocal}
\end{equation}
Here $\gamma_{I,l}$ is a weight factor, and each $b_{I,l}$ is a real
valued function. The function $b_{I,l}$ is also localized, in the sense that it is  compactly supported around $\vr=0$.
The locality originates from the physical meaning of nonlocal pseudopotentials, i.e. they characterize the orthogonality of the valence electron orbitals with respect to the core electron orbitals, and hence the support of $b_{I,l}$ is restricted by the span of the core orbitals. %In a pseudopotential approximation,  $V_I (\vr - \vR_I )$ is defined as
%\begin{equation}
%  V_{I}(\vr-\vR_{I}) := \int
%  v_{c}(\vr,\vr')m_{I}(\vr'-\vR_{I}) \ud \vr',
%  \label{eqn:pseudocharge}
%\end{equation}
%where $m_{I}$ is a localized function in the real space and is called
%a pseudocharge~\cite{Martin2004,PaskSterne2005}. The normalization condition for each pseudocharge is $\int m_I(\vr) \ud \vr = -Z_I$, 
%and $Z_I$ is the atomic charge for the $I$-th atom. The total pseudocharge is defined as
%$m(\vr) = \sum_{I} m_{I}(\vr-\vR_{I}).$
%We assume the system is charge neutral, i.e.
%\[
%\int m(\vr) \ud \vr = -\sum_I Z_I = -N_e.
%\]
$E_{\xc}$ is the exchange-correlation energy, and here we assume
semi-local functionals such as local
density approximation (LDA)~\cite{CeperleyAlder1980,PerdewZunger1981}
and generalized gradient approximation (GGA)
functionals~\cite{Becke1988,LeeYangParr1988,PerdewBurkeErnzerhof1996} are used.
$E_{\mathrm{II}}$ is the ion-ion Coulomb interaction energy. For
isolated clusters in 3D, 
$\displaystyle{ E_{\mathrm{II}}(\{\vR_{I}\}) = \frac{1}{2}\sum_{I\ne J}
\frac{Z_{I}Z_{J}}{\abs{\vR_{I}-\vR_{J}}} }$, while for periodic systems
the contribution from all the image charges should be properly taken
into account via e.g. the Ewald summation technique~\cite{FrenkelSmit2002}.  The
last term of Eq.~\eqref{eqn:KSfunc} is the entropy term related to the
temperature, and spin degeneracy is neglected for simplicity of the
notation. 

The Euler-Lagrange equation associated with the Kohn-Sham energy
functional gives rise to the Kohn-Sham equations as
\begin{align}
  & H[\rho] \psi_i = \left( -\frac{1}{2} \Delta +  \mcV[\rho] \right) \psi_i = \varepsilon_i \psi_i , \label{eqn:Hamil}\\
        & \int \psi_i^*(\vr) \psi_j ( \vr) \ud \vr = \delta_{ij}, \quad
  \rho(\vr) = \sum_{i=1}^{\infty} f_{i} \left| \psi_i(\vr) \right|^2, \quad
  f_{i} = \frac{1}{1+e^{\beta(\varepsilon_{i}-\mu)}}.
\end{align}
Here the eigenvalues $\{ \varepsilon_i \}$ are ordered non-decreasingly.
Note that
the occupation number $f_{i}$ is given analytically by the Fermi-Dirac
distribution with respect to the eigenvalue $\varepsilon_{i}$, and
$\mu$ is a Lagrange multiplier enforcing the normalization condition of
the electron density.
The difference of the eigenvalues
$\varepsilon_{g} = \varepsilon_{N_{e}+1} - \varepsilon_{N_{e}}$ is called
the energy gap. If $\varepsilon_{g}$ is positive, then the system is called an
insulating system. Otherwise it is a metallic system.
For insulating systems, 
$\psi_1,\ldots,\psi_{N_e}$ are called the occupied orbitals, while
$\psi_{N_e + 1},\ldots$ are called the unoccupied orbitals. 
$\psi_{N_{e}}$ is sometimes called the highest occupied
molecular orbital (HOMO), and $\psi_{N_{e}+1}$ the lowest
unoccupied molecular orbital (LUMO). 

The effective potential $\mcV[\rho]$ depends on the electron density
$\rho$ as
\begin{equation}
  \mcV [\rho](\vr,\vr') = V_{\mathrm{ion}}(\vr,\vr') + \left[\int v_c(\vr,\vr') \rho(\vr') \ud \vr' +
V_{\xc}[\rho](\vr)\right] \delta(\vr-\vr') .
    \label{eqn:effV}
\end{equation}
Here $\displaystyle{ V_{\xc}[\rho](\vr) = \frac{\delta E_{\xc}}{\delta \rho(\vr)} }$ is
the exchange-correlation potential, which is 
the functional derivative of the exchange-correlation energy with
respect to the electron density. The Kohn-Sham Hamiltonian depends nonlinearly on the electron density $\rho$, and the electron density
should be solved self-consistently.
When the Kohn-Sham energy functional $E_{\KS}$ achieves its
minimum, the self-consistency of the electron density is simultaneously
achieved. Note that both the Hartree potential and the
exchange-correlation potential are local potentials. This plays an
important role in simplifying the treatment of the density functional perturbation
theory.

%Then the total energy can be equivalently computed
%as~\cite{Martin2004}
%\REV{
%\begin{equation}
%       \begin{split}
%    E_{\mathrm{tot}} = &\sum_{i=1}^{N_{e}}\varepsilon_{i} - \frac{1}{2}\iint 
%     v_c(\vr,\vr')\rho(\vr)
%    \rho(\vr') \ud \vr \ud \vr'\\
%    &-\int V_{\xc}[\rho](\vr) \rho(\vr) \ud \vr + E_{\xc}[\rho] +
%    E_{\mathrm{II}}(\{\vR_{I}\}).
%    \end{split}
%    \label{eqn:Etotal}
%\end{equation}
%}
%Here $E_{\mathrm{band}} = \sum_{i=1}^{N_{e}}\varepsilon_{i}$
%is referred to as the band energy.

When the Kohn-Sham energy functional $E_{\KS}$ achieves its
minimum, the self-consistency of the electron density is simultaneously
achieved.  Then the total energy can be equivalently computed
as~\cite{Martin2004}
\begin{equation}
        \begin{split}
    E_{\text{tot}} = &\sum_{i=1}^{N_{e}}\varepsilon_{i} - \frac{1}{2}\iint 
     v_c(\vr,\vr')\rho(\vr)
    \rho(\vr') \ud \vr \ud \vr'\\
    &-\int V_{\xc}[\rho](\vr) \rho(\vr) \ud \vr + E_{\xc}[\rho] +
    E_{\mathrm{II}}(\{\vR_{I}\}).
    \end{split}
    \label{eqn:Etotal}
\end{equation}
Here $\displaystyle{ E_{\text{band}} = \sum_{i=1}^{N_{e}}\varepsilon_{i} }$
is referred to as the band energy.

At this point, the atomic force can be given by the negative of
the first order derivative of $E_{\mathrm{tot}}$ with
respect to the atomic configuration using the Hellmann-Feynman
theorem as
%\begin{equation}
%    \begin{split}
%    \vF_{I} =& -\frac{\partial E_{\mathrm{tot}}(\{\vR_{I}\})}{\partial
%    \vR_{I}} = -\int \frac{\partial
%    V_{\mathrm{ion}}}{\partial \vR_{I}}(\vr;\{\vR_{I}\}) \rho(\vr) \ud
%    \vr - \frac{\partial E_{\mathrm{II}}(\{\vR_{I}\})}{\partial \vR_{I}}\\
%    =&-\int \frac{\partial
%    V_{I}}{\partial \vR_{I}}(\vr-\vR_{I}) \rho(\vr) \ud
%    \vr - \frac{\partial E_{\mathrm{II}}(\{\vR_{I}\})}{\partial
%    \vR_{I}}.
%    \end{split}
%    \label{eqn:force}
%\end{equation} 
\begin{equation}
    \vF_{I} = -\frac{\partial E_{\mathrm{tot}}(\{\vR_{I}\})}{\partial
    \vR_{I}} = -\int \frac{\partial V_{\mathrm{ion}}}{\partial \vR_{I}}(\vr,\vr';\{\vR_{I}\})
    P(\vr',\vr) \ud
    \vr \ud \vr' - \frac{\partial E_{\mathrm{II}}(\{\vR_{I}\})}{\partial
    \vR_{I}}.
    \label{eqn:force}
\end{equation} 
Here $P$ is the density matrix defined as
\begin{equation}
  P(\vr,\vr') = \sum_{i=1}^{\infty} f_{i}
  \psi_{i}(\vr)\psi_{i}^{*}(\vr').
  \label{eqn:densitymatrix}
\end{equation}
In particular, the diagonal entries of the density matrix
$P(\vr,\vr)$ is the electron density $\rho(\vr)$.
The derivative of the pseudopotential $\displaystyle{\frac{\partial V_{\mathrm{ion}}}{\partial
\vR_{I}}(\vr,\vr';\{\vR_{I}\}) }$ does not depend on the electron density,
can be obtained semi-analytically. Hence the computation of the atomic
force only involves a number of quadratures.
The atomic force allows the performance of structural relaxation of the
atomic configuration, by minimizing the total energy $E_{\mathrm{tot}}$
with respect to the atomic positions $\{\vR_{I}\}$. 
When the atoms are at their equilibrium positions, all atomic forces should
be zero. 

\subsection{Density functional perturbation theory} \label{sec:DFPT}

In density functional perturbation theory (DFPT), we assume that the
self-consistent ground state electron density $\rho$ has been computed,
denoted by $\rho^{*}$.
In this paper, we focus on phonon calculations using DFPT.
Assume the system deviates from its equilibrium position $\{\vR_{I}\}$
by some small magnitude, then the changes of the total energy is
dominated by the Hessian matrix with respect to the atomic positions.
The dynamical matrix $D$ consists of 
$d\times d$ blocks in the form
\[
D_{I,J} = \frac{1}{\sqrt{M_{I}M_{J}}}\frac{\partial^2
E_{\mathrm{tot}}(\{\vR_{I}\})}{\partial \vR_{I} \partial \vR_{J}},
\]
where $M_{I}$ is the mass of the $I$-th nuclei. The dimension of the
dynamical matrix is $d\times N_{A}$. The equilibrium atomic
configuration is at a local minimum of the 
total energy, and all the
eigenvalues of $D$ are real and non-negative. Hence the eigen-decomposition of $D$ is
\[
D u_{k} = \omega_{k}^2 u_{k},
\]
where $u_{k}$ is called the $k$-th phonon mode, and $\omega_{k}$ is called the
$k$-th phonon frequency.
The phonon spectrum is defined as the distribution of the eigenvalues $\{\omega_k\}$ i.e.
\begin{equation}
\varrho_D(\omega)=\frac{1}{d N_A} \sum_{k} \delta(\omega-\omega_k).
\label{eqn:phononspec}
\end{equation}
Here $\delta$ is the Dirac-$\delta$ distribution. $\varrho_D$ is also referred to as the density of states of $D$~\cite{Martin2004,LinSaadYang2016}.

%In order to compute the Hessian matrix, we first define the density matrix as
%\begin{equation}
%    P[H](\vr,\vr') =
%    \sum_{i=1}^{N_{e}}\psi_{i}(\vr)\psi_{i}^{*}(\vr').
%\end{equation}
%It is clear that the electron density $\rho(\vr)=P[H](\vr,\vr)$ is the
%diagonal of the density matrix. Furthermore, when
%$\varepsilon_{g}>0$, the density matrix can be equivalently represented
%in the operator formulation as
%\begin{equation}
%    P[H] = f_{0}(H-\mu) = \frac{1}{2\pi i}\int_{\Gamma} (\lambda - H)^{-1} \ud \lambda.
%    \label{eqn:densitymat}
%\end{equation}
%Here $\mu$ is a real number between $\varepsilon_{N_{e}}$ and
%$\varepsilon_{N_{e}+1}$ and is called the chemical potential.
%$f_{0}(x)$ is a step function that takes the value $1$ if $x\le 0$ and
%$0$ if $x>0$.  The second equality in Eq.~\eqref{eqn:densitymat} is due
%to Cauchy's contour integral theorem, and $\Gamma$ is a contour
%encircles the first $N_e$ eigenvalues of $H$ and excluding others. 

In order to compute the Hessian matrix, we obtain from
Eq.~\eqref{eqn:force} that
\begin{equation}
    \begin{split}
        \frac{\partial^2 E_{\mathrm{tot}}(\{\vR_{I}\})}{\partial \bvec{R_{I}}
        \partial \bvec{R_{J}}}=& 
        \int \frac{\partial V_{\mathrm{ion}}}{\partial
        \vR_I}(\vr,\vr';\{\vR_{I}\}) \frac{\partial
        P(\vr',\vr)}{\partial \vR_J}
        \ud\vr \ud \vr' \\
        &+ \int \frac{\partial^2 V_{\mathrm{ion}}}{\partial
        \vR_I \partial \vR_{J}}(\vr,\vr';\{\vR_{I}\}) P(\vr',\vr) \ud
        \vr  \ud \vr' + \frac{\partial^2 E_{\mathrm{II}}(\{\vR_{I}\})}{\partial \vR_I
        \partial \vR_J}.\\
    \end{split}
    \label{eqn:SecDeriv}
\end{equation}

%\begin{align}
%\frac{\partial^2 E_{\mathrm{tot}}(\{\vR_{I}\})}{\partial \bvec{R_{I}}
%\partial \bvec{R_{J}}} & = \int \frac{\partial V_I}{\partial
%\vR_I}(\vr-\vR_{I}) \frac{\delta \rho(\vr)}{\delta \vR_J}
%\ud\vr + \int \rho(\vr) \frac{\partial^2 V_I}{\partial
%\vR_I^2}(\vr-\vR_{I}) \ud \vr \notag \\
%
%& = \int \frac{\partial V_I}{\partial \vR_I}(\vr-\vR_{I}) \frac{\delta
%\rho(\vr)}{\delta V(\vr')} \frac{\partial
%V_J}{\partial \vR_J}(\vr-\vR_{J}) \ud\vr' \ud\vr 
%+ \frac{\partial^2 E_{\mathrm{II}}(\{\vR_{I}\})}{\partial \vR_I \partial \vR_J}
%\\
%&+\delta_{\vR_{I},\vR_{J}}
%\int
%\rho(\vr) \frac{\partial^2 V_I}{\partial
%\vR_I^2}(\vr-\vR_{I})
%\ud \vr. \label{eqn:SecDeriv}
%\end{align}
%
%

Similar to the force calculation, the second term of
Eq.~\eqref{eqn:SecDeriv} can be readily computed with numerical
integration, and the third term involves only ion-ion interaction that
is independent of the electronic states. Hence the first term is the
most challenging one due to the response of the electron density with
respect to the perturbation of atomic positions. Applying the chain
rule, we have
\begin{equation}
  \begin{split}
        &\int \frac{\partial V_{\mathrm{ion}}}{\partial
        \vR_I}(\vr,\vr';\{\vR_{I}\}) \frac{\partial
        P(\vr',\vr)}{\partial \vR_J}
        \ud\vr \ud \vr' \\
    =&
    \int \frac{\partial V_{\mathrm{ion}}(\vr,\vr';\{\vR_{I}\})}{\partial
        \vR_I} \frac{\delta P(\vr',\vr)}{\delta
        V_{\mathrm{ion}}(\vr'',\vr''')} \frac{\partial
        V_{\mathrm{ion}}(\vr'',\vr''');\{\vR_{I}\})}{\partial \vR_{J}}
        \ud\vr \ud \vr' \ud \vr'' \ud \vr'''.
  \end{split}
  \label{eqn:chiterm}
\end{equation}
Here the Fr\'{e}chet derivative
$\displaystyle{\mf{X}(\vr,\vr';\vr'',\vr''') = \frac{\delta
P(\vr,\vr')}{\delta V_{\mathrm{ion}}(\vr'',\vr''')}}$ is referred to as
the reducible polarizability operator~\cite{OnidaReiningRubio2002}, which characterizes the
\textit{self-consistent} linear response of the density matrix at
$(\vr,\vr')$ with respect to an external nonlocal perturbation of $V_{\mathrm{ion}}$ at
$(\vr'',\vr''')$. However, the computation of
$\mf{X}$ must be obtained through a simpler quantity 
$\displaystyle{ \mf{X}_{0}(\vr,\vr';\vr'',\vr''') = \frac{\delta
P(\vr,\vr')}{\delta \mcV(\vr'',\vr''')} }$,
which is called the irreducible polarizability operator (a.k.a.
independent particle polarizability operator)~\cite{OnidaReiningRubio2002}. 

The discussion using the notation $\vr,\vr',\vr''$ etc will quickly
become complicated.  For simplicity in the discussion below, we will not distinguish the
continuous and discretized representations of various quantities. In
the case when a discretized representation is needed, we assume that the
computational domain is uniformly discretized into a number of grid
points $\{\vr_\alpha\}_{\alpha=1}^{N_g}$. After discretization all
quantities can be called tensors. For example,
we will call 
$u(\vr)$ an order $1$ tensor (or a vector), $A(\vr,\vr')$ an order $2$
tensor (or a matrix),
and $\mf{X}(\vr,\vr';\vr'',\vr''')$ an order $4$ tensor. The
tensor slicing and tensor contraction can be denoted using either the continuous or the
discrete notation. For example, $\mf{X}(\vr,\vr;\vr'',\vr''')$
denotes a sliced tensor which is an order $3$ tensor. 
The tensor contraction between two order $1$ tensors $u$ and $v$ should
be interpreted as $u^{*}v = \int u^{*}(\vr)v(\vr) \ud \vr$.
The tensor contraction between an order $2$ tensor $A$ and an order
$1$ tensor $v$ (i.e. a matrix-vector product) should be interpreted as
$(Av)(\vr) = \int A(\vr,\vr') v(\vr') \ud \vr'$.
Similarly the contraction between an order $2$ tensor $A$ and an order
$2$ tensor $\mf{g}$ (i.e. matrix-matrix product) 
should be interpreted as 
$(A\mf{g})(\vr,\vr') = \int A(\vr,\vr'') \mf{g}(\vr'',\vr')
\ud \vr''$, and the
contraction between an order $4$ tensor $\mf{X}$ and an order
$2$ tensor $\mf{g}$ 
should be interpreted as 
\[
(\mf{X}\mf{g})(\vr,\vr') = \int
\mf{X}(\vr,\vr';\vr'',\vr''') \mf{g}(\vr'',\vr''') \ud \vr''
\ud \vr'''.
\]
We also define two operations for order $1$ tensors. 
The Hadamard product of two order $1$ tensors $u\odot v$ should be interpreted
as $(u\odot v)(\vr) := u(\vr) v(\vr)$. For an order $1$ tensor $v(\vr)$, we
define an associated order $2$ tensor as $(\mathrm{diag}[v])(\vr,\vr') :=
v(\vr)\delta(\vr-\vr')$. It is easy to verify that the Hadamard product  can be written as $u\odot v =\mathrm{diag}[u] v$.

Using the linear algebra type of notation as above, the key difficulty
of phonon calculations
is the computation of the tensor contraction $\mf{u}=\mf{X}\mf{g}$, where $\mf{g}$
traverses $d\times N_{A}$ order $2$ tensors of the form $\displaystyle{ \frac{\partial
V_{\mathrm{ion}}(\vr'',\vr''';\{\vR_{I}\})}{\partial \vR_{J,a}} }$, where
$\vR_{J,a}$ is the $a$-th direction of the atomic position
$\vR_{J}$ ($a=1,\ldots,d$). According
to Eq.~\eqref{eqn:pseudopot}, $\mf{g}$ can split into a local
component and a nonlocal component as
\begin{equation}
  \mf{g}(\vr,\vr') = g_{\mathrm{loc}}(\vr) \delta(\vr-\vr') +
  \mf{g}_{\mathrm{nl}}(\vr,\vr'),
  \label{}
\end{equation} 
or equivalently $\mf{g} = \mathrm{diag}[g_{\mathrm{loc}}] +
\mf{g}_{\mathrm{nl}}$.
For each $\mf{g}$, only one atom $J$ contributes to
the order $1$ tensor $g_{\mathrm{loc}}$ and the order $2$ tensor
$\mf{g}_{\mathrm{nl}}$. From the definition of nonlocal pseodopotential Eq.~\eqref{eqn:nonlocal}, we have
\begin{equation}
\begin{split}
    & \mf{g}_{\mathrm{nl},I}(\vr, \vr') = \sum_{l=1} ^ {L_I} \gamma_{I,l} \left[ b_{I,l}(\vr - \vR_I) db^*_{I,l} (\vr'-\vR_I) + db_{I,l} (\vr - \vR_I) b_{I,l}^*(\vr' - \vR_I) \right],
    \label{eqn:gnl} \\ 
    & \text{where}  \quad db_{I, l}(\vr - \vR_I) := \frac{\partial b_{I,l}(\vr - \vR_I)}{\partial \vR_I}.
\end{split}
\end{equation}
We note that 
$\mf{g}_{\mathrm{nl}}$ is a symmetric operator of rank
$2L_{I}$, where the factor $2$ comes from
the Leibniz formula. In the rest of the paper, we shall use $b_{l}(\vr), db_{l}(\vr)$ to hide the explicit dependence on the atom indices $I$ or the atomic positions $\{\vR_I\}$.

%We may refer to quantities
%such as $U(\vr)\equiv [u_1(\vr),\ldots,u_{N_e}(\vr)]$ as a matrix of
%dimension $N_g\times N_e$. In particular, all indices in the subscript
%are interpreted as the column indices of the matrix, and row indices are
%given in the parenthesis if necessary. For example, $u_i$ refers to the
%$i$-th column of $U$, and $u_i(\vr_\alpha)$ refers to the matrix element
%with row index $\vr_\alpha$. We denote by $M_{ij}$ a vector with a
%stacked column index $ij$, which refers to the $(i+(j-1)N_1)$-th column
%of the matrix $M$. Here the index $i$ ranges from $1$ to $N_1$, and $j$
%from $1$ to $N_2$, respectively.

% Again using linear algebra notation, we can write  $u\odot v$ as $\mathrm{diag}(u) v$.
% Such linear algebra notation gives a unified way to describe the algorithms in the continuous and discrete formulation.
%  We will specify the discretization used in our numerical simulation in section~\ref{sec:numer}.

From the definition of $\mcV$ in
Eq.~\eqref{eqn:effV}, we apply the chain rule and have
%\begin{equation}
% \frac{\delta \mcV(\vr,\vr')}{\delta V_{\mathrm{ion}}(\vr'',\vr''')} =
% \delta(\vr-\vr'')\delta(\vr'-\vr''') + \int
% \mf{f}_{\mathrm{hxc}}(\vr,\vr'''') 
% \frac{\delta P(\vr'''',\vr'''')}{\delta V_{\mathrm{ion}}(\vr'',\vr''')}
% \ud \vr'''' \delta(\vr-\vr').
%  \label{}
%\end{equation}
\begin{align}
%    \chi = \frac{\delta \rho}{\delta \mcV} \left(I + \frac{\delta
%    \mcV}{\delta \rho} \frac{\delta \rho}{\delta V_{\mathrm{ion}}}\right) =
%    \chi_0 ( I + v_{\mathrm{hxc}} \chi), \label{eqn:chitmp}
\mf{u} = \mf{X}\mf{g} = \frac{\delta P}{\delta \mcV} \frac{\delta
\mcV}{\delta V_{\mathrm{ion}}} \mf{g} =
\mf{X}_{0} \mf{g} + \mf{X}_{0} \mf{f}_{\mathrm{hxc}} \mf{X} \mf{g} = 
\mf{X}_{0} \mf{g} + \mf{X}_{0} \mf{f}_{\mathrm{hxc}} \mf{u}.
\label{eqn:dyson}
\end{align}
In Eq.~\eqref{eqn:dyson},
\begin{equation}
  \begin{split}
  \mf{f}_{\mathrm{hxc}}(\vr,\vr';\vr'',\vr''') = &\left(v_{c}(\vr,\vr'') +
  \frac{\delta V_{\mathrm{xc}}[\rho^{*}](\vr)}{\delta \rho(\vr'')}  \right)
  \delta(\vr-\vr')\delta(\vr''-\vr''')\\
  := &f_{\mathrm{hxc}}(\vr,\vr'')\delta(\vr-\vr')\delta(\vr''-\vr''')
  \end{split}
  \label{eqn:fhxc}
\end{equation}
is an order $4$ tensor, which is the kernel characterizing the dependence of the $\mc{V}$ with respect
to the density matrix $P$ in the linear regime. 
Here $\frac{\delta
V_{\mathrm{xc}}[\rho^{*}](\vr)}{\delta \rho(\vr')}$ is called the
exchange-correlation kernel, which is a local operator in the LDA and
GGA formulations of the exchange-correlation functionals.
Therefore in Eq. \eqref{eqn:fhxc}, $\delta(\vr-\vr')$ comes from that the Hartree and exchange-correlation potentials are local, while $\delta(\vr''-\vr''')$ comes from that the nonlinear term only depends on the electron density, i.e. the diagonal elements of the density matrix. 
Eq.~\eqref{eqn:dyson} is called the \textit{Dyson equation}, and the solution
$\mf{u}$ should be solved self-consistently.

In order to solve the Dyson equation~\eqref{eqn:dyson}, we need to apply
$\mf{X}_0$ to order $2$ tensors of the form $\mf{g}$ or
$\mf{f}_{\mathrm{hxc}} \mf{u}$. By means of the eigenfunctions
$\psi_{i}$, the eigenvalues $\varepsilon_{i}$, and the occupation
numbers $f_{i}$,
$\mf{X}_{0}\mf{g}$ can be expressed using the Adler-Wiser
formula~\cite{Adler1962,Wiser1963}
\begin{align}
  (\mf{X}_{0}\mf{g})(\vr,\vr') = \sum_{i,a=1}^{\infty}
  \frac{f_{a}-f_{i}}{\varepsilon_{a}-\varepsilon_{i}}
  \psi_{a}(\vr)\left(\int \psi_{a}^{*}(\vr'') \mf{g}(\vr'',\vr''')
  \psi_{i}(\vr''') \ud \vr'' \ud \vr''' \right)\psi_{i}^{*}(\vr'),
  \label{eqn:AdlerWiserInt}
\end{align}
where the term when $i=a$ should be interpreted as the limit when
$\varepsilon_{a}\to \varepsilon_{i}$.
Using the linear algebra notation, Eq.~\eqref{eqn:AdlerWiserInt} can be
written as
\begin{align}
  \mf{X}_{0}\mf{g} = \sum_{i,a=1}^{\infty}
  \frac{f_{a}-f_{i}}{\varepsilon_{a}-\varepsilon_{i}}
  \psi_{a}(\psi_{a}^{*} \mf{g} \psi_{i})\psi_{i}^{*}. \label{eqn:AdlerWiser}
\end{align}
Since $\mf{g}$ is an Hermitian order $2$ tensor, $\mf{X}_{0}\mf{g}$
is also an Hermitian order $2$ tensor. If we truncate the infinite sum
in Eq.~\eqref{eqn:AdlerWiser} to a finite sum of states,
Eq.~\eqref{eqn:AdlerWiser} and Eq.~\eqref{eqn:dyson} can be solved
together to obtain $\mf{u}$, and therefore the Hessian
matrix~\eqref{eqn:SecDeriv} can be evaluated.

In order to observe the computational complexity of DFPT for phonon
calculations, let us first neglect the nonlocal pseudopotential
$V_{\mathrm{nl},I}$, which simplifies the discussion.
Since each $\mf{g}$ only involves the local contribution,
Eq.~\eqref{eqn:SecDeriv} only requires $\frac{\partial
\rho(\vr)}{\partial \vR_{J}}$. Therefore one is only interested in
computing
\begin{equation}
  u(\vr) = \mf{u}(\vr,\vr) = \int \mf{X}(\vr,\vr;\vr',\vr')
  \mf{g}(\vr',\vr') \ud \vr':= \int \chi(\vr,\vr')
  g_{\mathrm{loc}}(\vr') \ud \vr'.
  \label{eqn:twopointchi}
\end{equation}
Here we have introduced the notation $\chi(\vr,\vr') = \mf{X}(\vr,\vr;\vr',\vr')$,
and used that the nonlocal component of $\mf{g}$ vanishes. Similarly we
can define $\chi_{0}(\vr,\vr') = \mf{X}_{0}(\vr,\vr;\vr',\vr')$. We also
consider insulating systems with a finite band gap. This allows us to
reduce the temperature dependence of the occupation number, so that
$f_{i}=1$ if $i\le N_{e}$ and $0$ if $i\ge N_{e}+1$. As a result,
Eq.~\eqref{eqn:AdlerWiser} can be simplified as
\begin{equation}
  \chi_{0} g_{\mathrm{loc}} = \sum_{i=1}^{N_{e}}
  \sum_{a=N_{e}+1}^{\infty}
  \frac{1}{\varepsilon_{i}-\varepsilon_{a}} 
  \mathrm{diag}[\psi_{i}^{*}]
  \psi_{a}
  \left(\psi_{a}^{*}
  \mathrm{diag}[g_{\mathrm{loc}}]\psi_{i}\right) +
  \mathrm{h.c.}
  \label{eqn:AdlerWisertwopoint}
\end{equation}
Here $\mathrm{h.c.}$ means the Hermitian conjugate of the first term.

In order to overcome the difficulty of explicitly computing all the
unoccupied orbitals $\{\psi_{a}\}_{a=N_{e}+1}^{\infty}$, we first define
the projection operator to the unoccupied space 
$Q=I-\sum_{i=1}^{N_e} \psi_i\psi_i^*$.
Then we can compute $\chi_0 g_{\mathrm{loc}}$ as
\begin{equation}
\begin{split}
  \chi_0 g_{\mathrm{loc}} &=  \sum_{i=1}^{N_e} \mathrm{diag}[\psi_{i}^{*}] 
  Q (\varepsilon_{i}-H)^{-1} Q
  (\mathrm{diag}[g_{\mathrm{loc}}]\psi_{i}) + \mathrm{h.c.}.
\end{split}
\label{eqn:nounoccupy}
\end{equation}
In principle, since $Q$ commutes with $H$, the right hand side of Eq.~\eqref{eqn:nounoccupy}
only requires one $Q$ operator to be present. However, we choose the form
$Q (\varepsilon_i - H)^{-1} Q$ to emphasize that this operator is
Hermitian. Let $\zeta_i := Q (\varepsilon_i - H)^{-1} Q (
\mathrm{diag}[g_{\mathrm{loc}}] \psi_{i})$, the matrix inverse in
Eq.~\eqref{eqn:nounoccupy} can be avoided by solving the \textit{Sternheimer
equations}
\begin{equation}
  Q(\varepsilon_i - H)Q \zeta_i =Q(\mathrm{diag}[g_{\mathrm{loc}}]\psi_{i}).
  \label{eqn:Sternheimer}
\end{equation}
This strategy has been used in a number of contexts involving the
polarizability
operator~\cite{GonzeLee1997,OnidaReiningRubio2002,UmariStenuitBaroni2010,GiustinoCohenLouie2010,NguyenPhamRoccaEtAl2012}.
The Sternheimer equations can be solved using standard direct or iterative linear solvers. The choice of the solver can depend on practical matters such as the discretization scheme, and the availability of preconditioners. In practice for planewave discretization, we find that the use of the minimal residual method (MINRES)~\cite{PaigeSaunders1975} gives the best numerical performance.

The complexity of phonon calculations can now be analyzed as below. Even
with local pseudopotential only, and assume the Dyson equations always
converge within a constant number of iterations that is independent of
the system size $N_e$,
we need to apply $\chi_{0}$ to
$d\times N_{A}\sim \Or(N_{e})$ vectors of the form $g_{\mathrm{loc}}$.
Each $g_{\mathrm{loc}}$ requires solving $N_{e}$ Sternheimer
equations~\eqref{eqn:Sternheimer}, and the computational cost 
of applying the projection operator $Q$ to a vector is $\Or(N_e^2)$.
Hence the overall complexity is
$\Or(N_e^4)$~\cite{BaroniGironcoliDalEtAl2001}. This is significantly
more expensive than solving the KSDFT, of which the computational
complexity is typically $\Or(N_{e}^{3})$.

\subsection{Adaptively compressed polarizability operator}\label{subsec:acp}

%The recently developed adaptively compressed polarizability operator
%(ACP) formulation reduces the computational
%complexity of DFPT from $\Or(N_e^4)$ to $\Or(N_e^3)$ for the first time.
In this section we briefly review the ACP formulation~\cite{LinXuYing2017}
in the context of phonon calculations for insulating systems using local
pseudopotentials. If we label the possible $g_{\mathrm{loc}}$ using
a single index $j$, the Sternheimer equation~\eqref{eqn:Sternheimer} can be written as
\begin{equation}
  Q(\varepsilon_i - H)Q \zeta_{ij} =Q(\psi_{i} \odot g_{\mathrm{loc},j}).
  \label{eqn:eqnU}
\end{equation}
Here we have used the relation $\diag[g_{\mathrm{loc}}]\psi =
\psi\odot g_{\mathrm{loc}}$ to place $g_{\mathrm{loc}}$ and $\psi$ on a
more symmetric footing.
Then reduction of the computational complexity is achieved by means of
reducing the $\Or(N_e^2)$ equations in Eq.~\eqref{eqn:eqnU} to $\Or(N_e)$ equations with
systematic control of the accuracy. 

The compression of the right hand side vectors is performed via the
interpolative separable density
fitting method by Lu and Ying~\cite{LuYing2015}. 
Let us denote by $M$ the collection of right hand side vectors in
Eq.~\eqref{eqn:eqnU} without the $Q$ factor, i.e. $M_{ij} = \psi_i\odot
g_{\mathrm{loc},j}$.
Here we have used $ij$ as a stacked column index for the matrix $M$.
The dimension of $M$ is $N_g\times \Or(N_e^2)$. 
Due to the large number of columns of $M$, we seek for the following interpolative decomposition (ID) type of compression~\cite{ChengGimbutasMartinssonEtAl2005} for the matrix $M$, i.e.
\begin{equation}\label{eqn:MID}
M_{ij}(\vr) \approx \sum_{\mu=1}^{N_\mu} \xi_\mu(\vr) M_{ij}(\vr_\mu)
\equiv  \sum_{\mu=1}^{N_\mu} \xi_\mu(\vr) \psi_i(\vr_\mu)
g_{\mathrm{loc},j}(\vr_\mu).
\end{equation}
Here $\{\vr_\mu\}_{\mu=1}^{N_\mu}$ denotes a collection of selected row
indices (see Fig. 1 in~\cite{LinXuYing2017} for an illustration).  
Mathematically, the meaning of the indices $\{\vr_\mu\}$ is clear:
Eq.~\eqref{eqn:MID} simply states that
for any grid point $\vr$, the corresponding row vector $M_{:}(\vr)$ can be approximately expressed as the linear combination of the
selected rows $\{M_{:}(\vr_\mu)\}$. 
Since $N_{g}\sim N_{e}$, as $N_{e}$ increases, the column dimension
of $M$ (which is $\Or(N_{e}^2)$) can be larger than its row dimension
(which is $N_{g}$), and we can expect that the
vectors $\{\psi_{i}\odot g_{j}\}$ are approximately linearly dependent.
Such observation has been observed in the electronic structure
community under the name of density fitting or resolution of identity (RI)~\cite{Weigend2002,SodtSubotnikHead-Gordon2006,Foerster2008,UmariStenuitBaroni2009,RenRinkeBlumEtAl2012},
and the numerical rank of the matrix $M$ after truncation can be only
$\Or(N_{e})$ with a relatively small pre-constant.
This dimension reduction property has also been recently analyzed in \cite{LuSoggeSteinerberger}.
In the context of the interpolative
decomposition, our numerical results also
indicate that it is sufficient to choose
$N_\mu\sim \Or(N_e)$, and the pre-constant is small. 

%\begin{figure}[ht]
%  \begin{center}
%    {\includegraphics[width=0.6\textwidth]{isdf1.png}}
%  \end{center}
%  \caption{\REV{Interpolative decomposition of $M_{ij}(\vr)$.}}
%  \label{fig:isdf1}
%\end{figure}

One possible way of finding interpolative decomposition is to use a
pivoted QR factorization~\cite{ChanHansen1990,GuEisenstat1996}. However,
the computational complexity for compressing the dense matrix $M$ using
the interpolative decomposition is still $\Or(N_e^4)$.
The interpolative
separable density fitting method~\cite{LuYing2015} employs a two-step procedure to reduce this cost.
The first step is to use a 
fast down-sampling procedure, such as a subsampled random Fourier transform
(SRFT)~\cite{WoolfeLibertyRokhlinEtAl2008}, to transform the matrix $M$
into a matrix $\wt{M}$ of smaller dimension $N_g\times r N_e$, with $r$
a relatively small constant so that $r N_{e}$ is slightly larger than
$N_{\mu}$. The
second step is to apply the pivoted QR decomposition to $\wt{M}$
\begin{equation}
  \wt{M}^* \wt{\Pi} = \wt{Q}\wt{R},
  \label{eqn:QRMt}
\end{equation}
where $\wt{\Pi}$ is a permutation matrix and encodes the choice of the
row indices $\{\vr_{\mu}\}$ from $\wt{M}$. The interpolation vectors
$\{\xi_\mu\}$ in Eq.~\eqref{eqn:MID} can be also be computed from this
pivoted QR decomposition. It should be noted that the pre-processing
procedure does not affect the quality of the interpolative
decomposition, while the cost of the pivoted QR factorization in
Eq.~\eqref{eqn:QRMt} is now reduced to $\Or(N_{g} N_{\mu}^2) \sim
\Or(N_{e}^3)$.  We refer readers to~\cite{LuYing2015,LinXuYing2017} for
a more detailed description of this procedure. 

Once the compressed representation~\eqref{eqn:MID} is obtained, 
we solve the following set of modified Sternheimer equations
\[
Q(\varepsilon_i - H)Q \wt{\zeta}_{c\mu} = Q \xi_\mu,\quad i=1,\ldots,N_e,\quad \mu=1,\ldots,N_\mu.
\]
Note that there are still $\Or(N_{e}^2)$ equations to solve, but this
time the number of equations arises from the energy dependence on the
left hand side of the equation.  If the band gap is positive, we can
solve a set of equations of the form
\begin{equation}\label{eqn:eqncompressU}
Q(\wt{\varepsilon}_c - H)Q \wt{\zeta}_{c\mu} = Q \xi_\mu,\quad c=1,\ldots,N_c,\quad \mu=1,\ldots,N_\mu.
\end{equation}
where the number of shifts $N_{c}$ is independent of the system size
$N_{e}$. For example, this can be achieved using the Chebyshev points on
the occupied band $[\varepsilon_{1},\varepsilon_{N_{e}}]$, and the
number of Chebyshev points needed to achieve a certain error tolerance
scales weakly with respect to the band gap as
$\sqrt{|\mc{I}|/\varepsilon_{g}}$. Here $\varepsilon_{g}$ is the band
gap and $|\mc{I}|=\varepsilon_{N_{e}}-\varepsilon_{1}$ is the width of
the occupied band~\cite{LinXuYing2017}. 

Then define
\begin{equation}
  W_{\mu} = \sum_{i=1}^{N_e}  \mathrm{diag}[\psi_i^*] \odot \left(\sum_{c=1}^{N_c} \wt{\zeta}_{c \mu } \prod_{c'\ne c} \frac{\varepsilon_i -
\wt{\varepsilon}_{c'}}{\wt{\varepsilon}_c -
\wt{\varepsilon}_{c'}}\right) \psi_i(\vr_\mu)  + \mathrm{h.c.}, 
\label{eqn:eqnW}
\end{equation}
and we can combine Eq.~\eqref{eqn:eqnW} with
Eq.~\eqref{eqn:Sternheimer} to compute $\chi_0 g_{\mathrm{loc},j}$ as
\begin{equation}
\chi_0 g_{\mathrm{loc},j} \approx \sum_{\mu=1}^{N_\mu} W_\mu
g_{\mathrm{loc},j}(\vr_\mu).  \label{eqn:chi0gcompress}
\end{equation}

Formally, Eq.~\eqref{eqn:chi0gcompress} can further be simplified by
defining a matrix $\Pi$ with $N_\mu$ columns, which consists of selected
columns of a permutation matrix, i.e. 
$\Pi = \wt{\Pi}_{:,1:N_\mu}$ 
as the first $N_\mu$ columns of the permutation matrix obtained
from pivoted QR decomposition. More specifically,
$\Pi_{\mu}=e_{\vr_\mu}$ and $e_{\vr_\mu}$ is a unit vector with only one
nonzero entry at $\vr_\mu$ such that $e_{\vr_\mu}^T g_j=g_j(\vr_\mu)$.
Then 
\begin{equation}
\chi_0 g_{\mathrm{loc},j} \approx W \Pi^T g_{\mathrm{loc},j} := \wt{\chi}_0[\{g_{\mathrm{loc},j}\}] g_{\mathrm{loc},j}.
\label{eqn:chi0gcompress2}
\end{equation}
Note that the notation $\wt{\chi}_0[\{g_{\mathrm{loc},j}\}]$ emphasizes
the dependence on the vectors that $\wt{\chi}_0$ applies to.  In other words,
$\wt{\chi}_0[\{g_{\mathrm{loc},j}\}]$ is designed to only agree with $\chi_0$ when
applied to vectors $\{g_{\mathrm{loc},j}\}$, and the difference between
$\wt{\chi}_0$ and $\chi_0$ is not controlled in the space orthogonal to
that spanned by these vectors.  The rank of $\wt{\chi}_0[\{g_{\mathrm{loc},j}\}]$ is only $N_\mu$, while the singular values of $\chi_0$
have a much slower decay rate. 

In the case when only local pseudopotential is used, the Dyson
equation~\eqref{eqn:dyson} is simplified as
\begin{equation}
  u_{j} = \chi g_{\mathrm{loc},j} = u_{0,j} +
  \chi_{0} f_{\mathrm{hxc}} u_{j}.
  \label{eqn:dysonloc}
\end{equation}
Here $u_{0,j}:=\chi_{0} g_{\mathrm{loc},j}$ is called the
non-self-consistent response, and has been computed using 
the algorithm described above.

In order to solve Eq.~\eqref{eqn:dysonloc}, we do not only need
to evaluate $\chi_{0} g_{\mathrm{loc},j}$, but also the application of
$\chi_{0}$ to the self-consistent response $f_{\mathrm{hxc}} u_{j}$
which is not known \textit{a priori}. If we build a library of right hand
side vectors so that the application of $\chi_{0}$ remains accurate
throughout the iteration process of solving Eq.~\eqref{eqn:dysonloc},
the computational complexity can quickly increase. Instead it is much
more efficient to \textit{adaptively compress} the polarizability
operator $\chi_{0}$. 

Note that for any given set of functions $\{u_{j}\}$, we can construct
an operator $\wt{\chi}_{0}[\{f_{\mathrm{hxc}}u_{j}\}]$ so that $\wt{\chi}_{0}$ agrees well with
$\chi_{0}$ when applied to the vectors
$\{f_{\mathrm{hxc}}u_{j}\}$. The
Dyson equation can be rewritten as
\begin{equation}
  u_{j} = (I-\wt{\chi}_{0}[\{f_{\mathrm{hxc}}u_{j}\}])^{-1} u_{0,j}.
  \label{eqn:dysonloc2}
\end{equation}
Note that
$\wt{\chi}_{0}[\{f_{\mathrm{hxc}}u_{j}\}]$ is a low rank operator, and the matrix inverse in Eq.~\eqref{eqn:dysonloc2} can be efficiently
evaluated using the Sherman-Morrison-Woodbury formula. 

Eq.~\eqref{eqn:dysonloc2} yields an iterative scheme
\begin{equation}
  u^{k+1} = (I-\wt{\chi}_{0}[\{f_{\mathrm{hxc}}u^{k}\}])^{-1} u_0.
  \label{eqn:fixedpoint}
\end{equation}
In the equation we omitted the $j$ subindex of $u$. The convergence of the modified fixed point iteration \eqref{eqn:fixedpoint} can be understood as follows. 
At the iteration step $k$, the scheme and the true solution respectively satisfy 
\begin{equation}
\begin{split}
u^{k+1} &= u_0 + \wt{\chi}_{0}[\{f_{\mathrm{hxc}}u^{k}\}] f_{\mathrm{hxc}} u^{k+1}, \\
u^{*} &= u_0 + \chi_0 f_{\mathrm{hxc}} u^{*}. \\
\end{split}
\end{equation}
Let $e^{k} = u^{k} - u^*$ be the error at the iteration step $k$. We have 
\begin{equation}
\begin{split}
e^{k+1} &= \wt{\chi}_{0}[\{f_{\mathrm{hxc}}u^{k}\}] f_{\mathrm{hxc}} u^{k+1} - \chi_0 f_{\mathrm{hxc}} u^{*} \\ 
&= \wt{\chi}_{0}[\{f_{\mathrm{hxc}}u^{k}\}] f_{\mathrm{hxc}} u^{k+1} - \chi_{0} f_{\mathrm{hxc}} u^{k+1} + \chi_{0} f_{\mathrm{hxc}} u^{k+1}  - \chi_0 f_{\mathrm{hxc}} u^{*} \\
&= \eta^{k} + \chi_{0} f_{\mathrm{hxc}} e^{k+1}.
\end{split}
\end{equation}
Here 
\begin{equation}
\eta^{k} := ( \wt{\chi}_{0}[\{f_{\mathrm{hxc}}u^{k}\}] - \chi_0 ) f_{\mathrm{hxc}} u^{k+1},
\label{eqn:eta_chi0}
\end{equation}
which characterizes the discrepancy between $\wt{\chi}_{0}$ and $\chi_0$ when applied to the unknown vector $f_{\mathrm{hxc}} u^{k+1}$. 
Therefore the error at the $(k+1)$-th step satisfies 
\begin{equation}
\begin{split}
e^{k+1} &= (I - \chi_{0} f_{\mathrm{hxc}})^{-1} \eta^{k}.
\end{split}
\end{equation}
Since $\chi_{0}$ is negative semi-definite,  the norm of $(I - \chi_{0} f_{\mathrm{hxc}})^{-1}$ is bounded from above by one. Hence  the error goes to zero if the error of compression $\eta^{k}$ converges to $0$.

To summarize, the ACP formulation has three key ingredients: Compress
the right hand side; Disentangle the energy dependence; Adaptively
compress the polarizability operator. 
%We refer readers to~\cite{LinXuYing2017} for more details of the procedure.

\section{Split representation of the adaptively compressed
polarizability operator} \label{sec:splitACP}
%\ZX{Add a figure showing all interpolation points in split representation.}
In this section, we demonstrate how to generalize the ACP formulation in
section~\ref{subsec:acp} for efficient phonon calculations of real
materials. To this end we need to treat the nonlocal pseudopotential, as
well as temperature effects especially for metallic systems. We
demonstrate that the new 
split representation maintains the
$\Or(N_e^3)$ complexity, and improves all key steps in the ACP formulation, including Chebyshev interpolation of energy levels,
iterative solution of Sternheimer equations, and convergence of the Dyson equations.

The split representation of the polarizability operator first chooses
two cutoff energies
$\varepsilon_{\wt{N}_{\mathrm{cut}}}>\varepsilon_{N_{\mathrm{cut}}}\ge\mu$, and splits the right
hand side of Eq.~\eqref{eqn:AdlerWiser} into two terms

\begin{equation}
  \begin{split}
    \mf{X}_{0}\mf{g} \approx &\left[ \left( \sum_{i=1}^{N_{\mathrm{cut}}}  \sum_{a=N_{\mathrm{cut}} +1}^{\wt{N}_{\mathrm{cut}}}
  \frac{f_{a}-f_{i}}{\varepsilon_{a}-\varepsilon_{i}}
  \psi_{a}(\psi_{a}^{*} \mf{g} \psi_{i})\psi_{i}^{*} + \mathrm{h.c.} \right)  \right. \\
  &+\left. \sum_{i=1}^{N_{\mathrm{cut}}}  \sum_{a=1}^{N_{\mathrm{cut}}}
  \frac{f_{a}-f_{i}}{\varepsilon_{a}-\varepsilon_{i}}
  \psi_{a}(\psi_{a}^{*} \mf{g} \psi_{i})\psi_{i}^{*} \right] \\
  &+ 
  \left[\sum_{i=1}^{N_{\mathrm{cut}}}\sum_{a=\wt{N}_{\mathrm{cut}}+1}^{\infty}
  \frac{f_{i}}{\varepsilon_{i}-\varepsilon_{a}}
  \psi_{a}(\psi_{a}^{*} \mf{g} \psi_{i})\psi_{i}^{*} +
  \mathrm{h.c.}\right]\\
  :=& \mf{X}_{0}^{(s)} \mf{g} + \mf{X}_{0}^{(r)} \mf{g}.
  \end{split}
  \label{eqn:splitchi0}
\end{equation}
Here the first and second brackets split $\mf{X}_{0}\mf{g}$ into a
singular component $\mf{X}_{0}^{(s)} \mf{g}$ and a regular component 
$\mf{X}_{0}^{(r)} \mf{g}$, respectively. The Hermitian conjugate appears for the same
reason as in Eq.~\eqref{eqn:AdlerWisertwopoint} when treating insulating
systems.  $\mf{X}_{0}^{(s)}$ is called the singular component because for
systems with small gaps, the ratio
$(f_{a}-f_{i})/(\varepsilon_{a}-\varepsilon_{i})$ can be as large as
$1/\varepsilon_{g}$.  When the physical band gap $\varepsilon_{g}$ is small, this term becomes
numerically singular to treat in the iterative solution of Sternheimer
equations as well as the Chebyshev interpolation.  On the other hand,
the term $f_{i}/(\varepsilon_{i}-\varepsilon_{a})$ is bounded from above
by $1/\wt{\varepsilon}_{g}$, where $\wt{\varepsilon}_{g} = \varepsilon_{\wt{N}_{\text{cut}}+1}-\varepsilon_{N_{\text{cut}}}$ is called the \textit{effective gap}. As the effective gap
$\wt{\varepsilon}_{g}$ increases, the magnitude of $\mf{X}_{0}^{(r)}$
also decreases. In order to efficiently treat the singular part, we assume that the eigenfunctions $\{\psi_k\}_{k=1}^{\wt{N}_{\text{cut}}}$ have been computed using an iterative eigensolver. The cost for obtaining the additional eigenvectors is modest, given that the ground state DFT calculation already prepares the eigenvectors $\{\psi_k\}_{k=1}^{N_{\text{cut}}}$. 

\begin{figure}[ht]
    \centering
    \includegraphics[width=0.75\linewidth]{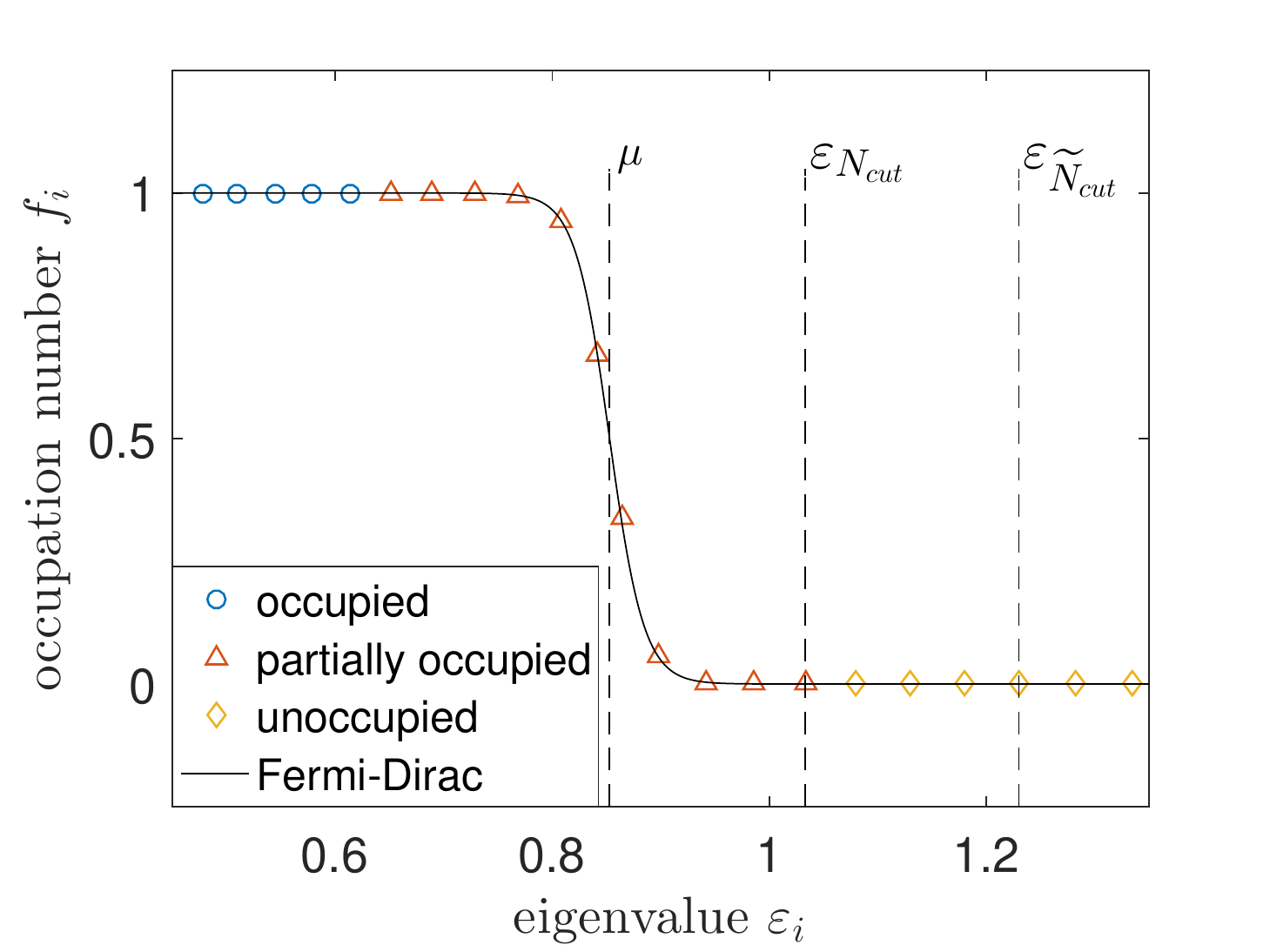}
    \caption{Schematic illustration of the cutoff energies with respect to the Fermi-Dirac distribution.}
    \label{fig:1Dground_cutoff}
\end{figure}

The approximation in
Eq.~\eqref{eqn:splitchi0} only comes from that as $\varepsilon$
increases above the chemical potential $\mu$,
the occupation number $f_{i} =
\frac{1}{1+e^{\beta(\varepsilon_{i}-\mu)}}$ decays exponentially. Then we can choose
$\varepsilon_{N_{\mathrm{cut}}}$ large enough so that
$f\left(\varepsilon_{N_{\mathrm{cut}}+1}\right)$ is sufficiently small and can
be approximated by $0$. For
insulating systems we can simply choose $N_{\mathrm{cut}}=N_{e}$.
The second energy cutoff $\varepsilon_{\wt{N}_{\mathrm{cut}}}$ defines an effective gap $\wt{\varepsilon}_{g}$,
of which the role will be discussed later. The split representation
requires the solution of eigenpairs $(\varepsilon_{i},\psi_{i})$ of $H$
for $i\le \wt{N}_{\mathrm{cut}}$. Fig. \ref{fig:1Dground_cutoff} illustrates the position of the cutoff energies along the energy spectrum, with respect to the occupation number given by the Fermi-Dirac distribution.

\subsection{Compression of the regular component of the
polarizability operator}
\label{sec:CompRegular}

One advantage of the split representation is that in the regular
component, the contribution from $f_{a}$ vanishes, and hence 
$\mf{X}_{0}^{(r)}\mf{g}$ can be evaluated using Sternheimer equations to
eliminate the need of computing all the unoccupied orbitals as follows
\begin{equation}
  \mf{X}_{0}^{(r)} \mf{g} = \sum_{i=1}^{N_{\mathrm{cut}}} f_{i} Q_{c}
  (\varepsilon_{i}-H)^{-1}Q_{c} (\mf{g} \psi_{i}) \psi_{i}^{*} +
  \mathrm{h.c.}
  \label{eqn:X0rg}
\end{equation}
Here the projection operator
$Q_{c}=I-\sum_{i=1}^{\wt{N}_{\mathrm{cut}}}\psi_{i}\psi_{i}^{*}$ projects a vector to the space which is orthogonal to the space 
 spanned by $\{\psi_{i}\}_{i=1}^{\wt{N}_{\mathrm{cut}}}$. When all
order $2$ tensors $\{\mf{g}_{j}\}$ are considered together,
Eq.~\eqref{eqn:X0rg} requires the solution of
\begin{equation}
  Q_{c}(\varepsilon_{i}-H) Q_{c} \zeta_{ij} = Q_{c}
  (\mf{g}_{j} \psi_{i}), \quad i=1,\ldots,N_{\mathrm{cut}},\quad j=1,\ldots,d\times
  N_A.
  \label{eqn:X0Sternhemier}
\end{equation}
Here each solution $\zeta_{ij}$ is still a vector.
The adaptive compression of $\mf{X}_{0} \mf{g}_{j}$ then parallels the
adaptive compression of $\chi_{0} g_{\mathrm{loc},j}$ as in
section~\ref{subsec:acp}, as detailed below.

The first step is to construct the collection of the right hand side
vectors $M_{ij} =
\mf{g}_{j} \psi_{i}$. Since the kernel of the nonlocal pseudopotential
from each atom is compactly supported in the real space, the
computational cost for generating $M$ is in fact dominated by the cost
associated with the local component $g_{\mathrm{loc},j}$. Hence the
overall cost is still $\Or(N_{e}^{3})$. The interpolative separable
density fitting procedure can then proceed as before, and generate a set
of compressed vectors $\{\xi_{\mu}\}_{\mu=1}^{N_{\mu}}$ as well as the
selected columns $\{\vr_{\mu}\}_{\mu=1}^{N_{\mu}}$.   The interpolation
decomposition then reads
\begin{equation}
  M_{ij}(\vr) = (\mf{g}_{j}\psi_{i})(\vr) \approx \sum_{\mu}
  \xi_{\mu}(\vr) (\mf{g}_{j}\psi_{i})(\vr_{\mu}).
  \label{eqn:X0interpdecompose}
\end{equation}

The second step is the disentanglement of the energy dependence. We
choose the Chebyshev interpolation points on the interval
$\mc{I} = [\varepsilon_{1},\varepsilon_{N_{\mathrm{cut}}}]$. 
Since the number of Chebyshev interpolation points is now controlled by
the effective gap
as $N_{c}\sim \Or(\sqrt{\mc{I}/\wt{\varepsilon}_{g}})$. Note that the gap $\epsilon_g$ (which can be small or zero) is now replaced by the effective gap $\wt{\varepsilon}_g$.  In practice we
observe that it is often sufficient to choose $N_{c}$ to be $5\sim 10$.

With the Chebyshev interpolation procedure, the Sternheimer equation
still takes the form~\eqref{eqn:eqncompressU}, with $Q$ replaced by
$Q_{c}$. The operator $Q_{c}(\varepsilon_{c}-H)Q_{c}$ is a negative
definite operator, with eigenvalue bounded from above by
$-\wt{\varepsilon_{g}}$. As the effective gap increases, the linear
system associated with the Sternheimer equation also becomes better
conditioned and the number of MINRES iterations can decrease. Typically we observe that MINRES can converge with around 10 steps. 

After the solution of the Sternheimer equations,
Eq.~\eqref{eqn:X0rg} becomes
\begin{equation}
  \mf{X}_{0}^{(r)} \mf{g}_{j} \approx
  \sum_{i=1}^{N_{\mathrm{cut}}}\sum_{\mu=1}^{N_{\mu}} f_{i} 
(\mf{g}_{j}\psi_i)(\vr_\mu) 
\left(\sum_{c=1}^{N_c} 
\wt{\zeta}_{c \mu } \prod_{c'\ne c} \frac{\varepsilon_i -
\wt{\varepsilon}_{c'}}{\wt{\varepsilon}_c - 
\wt{\varepsilon}_{c'}} \right) \psi_{i}^{*} +
  \mathrm{h.c.}
\label{eqn:X0rg2}
\end{equation}
Since that $\mf{g}_{j}$ can be split into a local and a nonlocal
component, we have
\begin{equation}
  (\mf{g}_{j}\psi_{i})(\vr_{\mu}) = g_{\mathrm{loc},j}(\vr_{\mu})
  \psi_{i}(\vr_{\mu}) + (\mf{g}_{\mathrm{nl},j} \psi_{i})(\vr_{\mu}).
  \label{}
\end{equation}
Define 
\begin{equation}
  \mf{W}^{(r)}_{\mu} = \sum_{i=1}^{N_{\mathrm{cut}}}
  \left(\sum_{c=1}^{N_c} \wt{\zeta}_{c \mu } \prod_{c'\ne c} \frac{\varepsilon_i -
\wt{\varepsilon}_{c'}}{\wt{\varepsilon}_c -
\wt{\varepsilon}_{c'}}\right) \psi_{i}(\vr_{\mu})f_{i} \psi_{i}^{*}  + \mathrm{h.c.},
  \label{eqn:Wr}
\end{equation}
and introduce the permutation matrix $\Pi$ as in
Eq.~\eqref{eqn:chi0gcompress2},
then Eq.~\eqref{eqn:X0rg2} becomes
\begin{equation}
  \begin{split}
  \mf{X}_{0}^{(r)} \mf{g}_{j} \approx&
  \sum_{\mu=1}^{N_\mu} \mf{W}^{(r)}_{\mu} (\Pi_{\mu}^{T} g_{\mathrm{loc},j}) 
  \\
  &+ \left[ \sum_{i=1}^{N_{\mathrm{cut}}}\sum_{\mu=1}^{N_{\mu}} f_{i} 
  (\mf{g}_{\mathrm{nl},j}\psi_i)(\vr_\mu) 
\left(\sum_{c=1}^{N_c} 
\wt{\zeta}_{c \mu } \prod_{c'\ne c} \frac{\varepsilon_i -
\wt{\varepsilon}_{c'}}{\wt{\varepsilon}_c -
\wt{\varepsilon}_{c'}} \right) \psi_{i}^{*} +
  \mathrm{h.c.} \right]
  \end{split}
  \label{eqn:X0rg3}
\end{equation}
At first glance, Eq.~\eqref{eqn:X0rg3} does not lead to any
simplification compared to Eq.~\eqref{eqn:X0rg2}. However, 
since the nonlocal component of $\mf{g}_{j}$ is compactly supported, for
each $\mf{g}_{\mathrm{nl},j}$ there are only $\Or(1)$ number of points
$\{\vr_{\mu}\}$ that contributes to $(\mf{g}_{\mathrm{nl},j}
\psi_{i})(\vr_{\mu})$. Hence the last term in Eq.~\eqref{eqn:X0rg3} is
much easier to evaluate than the direct evaluation of
Eq.~\eqref{eqn:X0rg2}.

\subsection{Compression of the singular component of the
polarizability operator}
\label{sec:CompSingular}

In practical calculations, numerical results indicate that it can be sufficient to choose
$\wt{N}_{\mathrm{cut}}\le 2N_{e}$, and hence the computation of
$\mf{X}^{(s)}_{0}\mf{g}$ can even be directly evaluated according to
Eq.~\eqref{eqn:splitchi0}. Compared to Eq.~\eqref{eqn:AdlerWiser}, the
computation of $\mf{X}^{(s)}_{0}\mf{g}$ still scales as
$\Or(N_{e}^{4})$, but the preconstant is much smaller.  In this section
we demonstrate that with a contour integral reformulation, we can
compress the singular component as well with $\Or(N_{e}^{3})$
complexity.

According to the derivation in Appendix A, $\mf{X}_{0}^{(s)}
\mf{g}$ can be evaluated using the contour integral formulation as
\begin{equation}
        \begin{split}
        \mf{X}_{0}^{(s)} \mf{g} =& \left[ \frac{1}{2\pi \I} \oint_{\mc{C}} f(z)
        (z-H_{c,2})^{-1}\mf{g}(z-H_{c,1})^{-1} \ud z  + \mathrm{h.c.} \right] \\
         &+ \frac{1}{2\pi \I} \oint_{\mc{C}} f(z)
        (z-H_{c,1})^{-1}\mf{g}(z-H_{c,1})^{-1} \ud z
        \end{split}
  \label{eqn:contourX0s}
\end{equation}
Here $H_{c,1}=\sum_{i=1}^{N_{\mathrm{cut}}} \psi_{i}
\varepsilon_{i}\psi_{i}^{*}, H_{c,2}=\sum_{i=N_{\mathrm{cut}}+1}^{\wt{N}_{\mathrm{cut}}} \psi_{i}
\varepsilon_{i}\psi_{i}^{*}$ are the Hamiltonian operators projected to
the subspace spanned by the first $N_{\mathrm{cut}}$ states, and to
the subspace spanned by the following $(\wt{N}_{\mathrm{cut}} - N_{\mathrm{cut}})$ states, respectively.
Before moving on to further discussion, we note that the numerically exact spectral decomposition of $H_{c,1}$ and $H_{c,2}$ is the key to reducing the complexity.

The contour integral in Eq. \eqref{eqn:contourX0s} can be discretized to obtain a numerical scheme. Let the integration nodes and weights be denoted by $\{z_p,\omega_p\}_{p=1}^{N_p}$, i.e. 
\begin{equation}
\frac{1}{2\pi \I} \oint_{\mc{C}} h(z)\ud z\approx   \sum_{p=1}^{N_p} \omega_{p} h(z_{p}),
\label{eqn:pole}
\end{equation}
for suitable $h(z)$, and the discretization scheme can be obtained using rational approximation methods~\cite{LinLuYingE2009,AAATrefethen,Moussa2016}. Then we have
\begin{equation}
  \begin{split}
    \mf{X}_{0}^{(s)} \mf{g}_{j} \approx & \left[ \sum_{p=1}^{N_p} \omega_{p}
    (z_{p}-H_{c,2})^{-1} \mf{g}_{j} (z_{p}-H_{c,1})^{-1} + \mathrm{h.c.} \right]\\
    &+ \sum_{p=1}^{N_p} \omega_{p}
    (z_{p}-H_{c,1})^{-1} \mf{g}_{j} (z_{p}-H_{c,1})^{-1} \\
  = & \left[ \sum_{p=1}^{N_p} \omega_{p} \sum_{i=1}^{N_{\mathrm{cut}}}
  (z_{p}-H_{c,2})^{-1}(\mf{g}_{j} \psi_{i})
  (z_{p}-\varepsilon_{i})^{-1}\psi_{i}^{*} + \mathrm{h.c.} \right] \\
  &+ \sum_{p=1}^{N_p} \omega_{p} \sum_{i=1}^{N_{\mathrm{cut}}}
  (z_{p}-H_{c,1})^{-1}(\mf{g}_{j} \psi_{i})
  (z_{p}-\varepsilon_{i})^{-1}\psi_{i}^{*},
  \end{split}
  \label{eqn:contourX0sPole}
\end{equation}
where the equality is derived from the spectral decompositions of
$H_{c,1},H_{c,2}$. When all $\{\mf{g}_{j}\}$ are considered together, we use again
the interpolative separable density
fitting~\eqref{eqn:X0interpdecompose} and obtain
\begin{equation}
  \begin{split}
  \mf{X}_{0}^{(s)} \mf{g}_{j} \approx & 
  \left[ \sum_{p=1}^{N_p} \omega_{p} \sum_{i=1}^{N_{\mathrm{cut}}}
  (z_{p}-H_{c,2})^{-1}\sum_{\mu=1}^{N_{\mu}}\xi_{\mu} (\mf{g}_{j}
  \psi_{i})(\vr_{\mu})
  (z_{p}-\varepsilon_{i})^{-1}\psi_{i}^{*} + \mathrm{h.c.} \right] \\
  &+ \sum_{p=1}^{N_p} \omega_{p} \sum_{i=1}^{N_{\mathrm{cut}}}
  (z_{p}-H_{c,1})^{-1}\sum_{\mu=1}^{N_{\mu}}\xi_{\mu} (\mf{g}_{j}
  \psi_{i})(\vr_{\mu}) 
  (z_{p}-\varepsilon_{i})^{-1}\psi_{i}^{*}  \\
  =& \left[ \sum_{i=1}^{N_{\mathrm{cut}}} \sum_{\mu=1}^{N_{\mu}}
  (\mf{g}_{j} \psi_{i})(\vr_{\mu}) \left( \sum_{p=1}^{N_{p}}
  \wt{\zeta}^{(s)}_{2,p\mu} \omega_{p} (z_{p}-\varepsilon_{i})^{-1}
  \right) \psi_{i}^{*}  + \mathrm{h.c.} \right] \\
  &+ \sum_{i=1}^{N_{\mathrm{cut}}} \sum_{\mu=1}^{N_{\mu}}
  (\mf{g}_{j} \psi_{i})(\vr_{\mu}) \left( \sum_{p=1}^{N_{p}}
  \wt{\zeta}^{(s)}_{1,p\mu} \omega_{p} (z_{p}-\varepsilon_{i})^{-1}
  \right) \psi_{i}^{*}.
  \end{split}
  \label{eqn:X0sg}
\end{equation}
In the last equation of~\eqref{eqn:X0sg}, we have defined the
solution $\wt{\zeta}^{(s)}_{\theta,p\mu}:=(z_{p}-H_{c,\theta})^{-1}\xi_{\mu}, \theta = 1,2$, which
can be numerically exactly computed from the spectral decompositions of
$H_{c,1},H_{c,2}$ respectively. We use the same strategy as in Eq.~\eqref{eqn:X0rg3} to handle
the contribution from $(\mf{g}_{j} \psi_{i})(\vr_{\mu})$. Define 
\begin{equation}
\begin{split}
\mf{W}^{(s)}_{\mu} =& \left[ \sum_{i=1}^{N_{\mathrm{cut}}} 
\psi_{i}(\vr_{\mu}) \left( \sum_{p=1}^{N_{p}}
\wt{\zeta}^{(s)}_{2,p\mu} \omega_{p} (z_{p}-\varepsilon_{i})^{-1}
\right) \psi_{i}^{*}  + \mathrm{h.c.} \right] \\
&+  \sum_{i=1}^{N_{\mathrm{cut}}} 
\psi_{i}(\vr_{\mu}) \left( \sum_{p=1}^{N_{p}}
\wt{\zeta}^{(s)}_{1,p\mu} \omega_{p} (z_{p}-\varepsilon_{i})^{-1}
\right) \psi_{i}^{*},
\end{split}
\label{eqn:Ws}
\end{equation}
and use the same permutation matrix $\Pi$ as in
Eq.~\eqref{eqn:chi0gcompress2}, then Eq.~\eqref{eqn:X0sg} becomes
\begin{equation}
\begin{split}
\mf{X}_{0}^{(s)} \mf{g}_{j} \approx&
\sum_{\mu=1}^{N_{\mu}} \mf{W}^{(s)}_{\mu} (\Pi_\mu^{T} g_{\mathrm{loc},j}) 
\\
&+ \left[ \sum_{i=1}^{N_{\mathrm{cut}}} \sum_{\mu=1}^{N_{\mu}}
(\mf{g}_{\mathrm{nl},j} \psi_{i})(\vr_{\mu}) \left( \sum_{p=1}^{N_{p}}
\wt{\zeta}^{(s)}_{2,p\mu} \omega_{p} (z_{p}-\varepsilon_{i})^{-1}
\right) \psi_{i}^{*}  + \mathrm{h.c.} \right]\\
&+ \sum_{i=1}^{N_{\mathrm{cut}}} \sum_{\mu=1}^{N_{\mu}}
(\mf{g}_{\mathrm{nl},j} \psi_{i})(\vr_{\mu}) \left( \sum_{p=1}^{N_{p}}
\wt{\zeta}^{(s)}_{1,p\mu} \omega_{p} (z_{p}-\varepsilon_{i})^{-1}
\right) \psi_{i}^{*}.
\end{split}
\label{eqn:X0sg2}
\end{equation}

%The analytic structure of the Fermi-Dirac function is shown in
%Fig.~\ref{fig:fermidirac}. If we choose 

%\LL{Add a picture of the contour. Check whether arbitrary contour integral discretization would work here for the singular part.}

\subsection{Adaptive compression for solving the Dyson equation}

Recall the Dyson equation~\eqref{eqn:dyson}, and so far we have computed
the non-self-consistent response
$\mf{u}_{0,j}:=\mf{X}_{0}\mf{g}_{j}$ using the split representation. In
order to solve the Dyson equation, we still need to evaluate
$\mf{X}_{0}\mf{f}_{\mathrm{hxc}}\mf{u}$ self-consistently. Use the
locality structure of $\mf{f}_{\mathrm{hxc}}$ as in
Eq.~\eqref{eqn:fhxc}, we have
\begin{equation}
  (\mf{X}_{0}\mf{f}_{\mathrm{hxc}}\mf{u})(\vr,\vr') = 
  \int \mf{X}_{0}(\vr,\vr';\vr'',\vr'') f_{\mathrm{hxc}}(\vr'',\vr''')
  \mf{u}(\vr''',\vr''') \ud \vr'' \ud \vr'''.
  \label{eqn:X0fu}
\end{equation}
It is important to observe that Eq.~\eqref{eqn:X0fu} only requires the
\textit{diagonal elements} of $\mf{u}$. Hence the self-consistent
solution of the Dyson equation~\eqref{eqn:dyson} only requires a set of
equations for these diagonal elements:
\begin{equation}
  \mf{u}_{j}(\vr,\vr) = \mf{u}_{0,j}(\vr,\vr) + 
  \int \mf{X}_{0}(\vr,\vr;\vr'',\vr'') f_{\mathrm{hxc}}(\vr'',\vr''')
  \mf{u}(\vr''',\vr''') \ud \vr'' \ud \vr'''.
  \label{eqn:dysonX02}
\end{equation}
Define $u_{j}(\vr)=\mf{u}_{j}(\vr,\vr)$ and
$u_{0,j}(\vr)=\mf{u}_{0,j}(\vr,\vr)$ and use the linear algebra
notation, then Eq.~\eqref{eqn:dysonX02} becomes a reduced Dyson equation
\begin{equation}
  u_{j} = u_{0,j} + \chi_{0} f_{\mathrm{hxc}} u_{j}.
  \label{eqn:dysonX03}
\end{equation}
Note that Eq.~\eqref{eqn:dysonX03} becomes precisely
the same as Eq.~\eqref{eqn:dysonloc}, which does not involve nonlocal pseudopotentials.
However, the important difference is that in Eq.~\eqref{eqn:dysonX03},   $u_{0,j}$ is taken from the
diagonal elements of $\mf{u}_{0,j}$, which properly takes into account
the nonlocal pseudopotential both in the Hamiltonian and in the
non-self-consistent response. 

Before moving on to the discussion of solving the reduced Dyson equation, we write out the explicit format of the diagonal part $u_{0,j} = \mf{u}_{0,j}$. Define $W_\mu^{(r)}(\vr) = \mf{W}_\mu^{(r)}(\vr, \vr), W_\mu^{(s)}(\vr) = \mf{W}_\mu^{(s)}(\vr, \vr)$, the diagonal part of Eq.~\eqref{eqn:X0rg3} reads
\begin{equation}
  \begin{split}
  \left( \mf{X}_{0}^{(r)} \mf{g}_{j}\right)(\vr,\vr)  \approx& \sum_{\mu=1}^{N_\mu} W^{(r)}_{\mu} (\vr) (\Pi_{\mu}^{T} g_{\mathrm{loc},j}) 
  \\
  &+ \left[  \sum_{i=1}^{N_{\mathrm{cut}}}\sum_{\mu=1}^{N_{\mu}} f_{i} 
  (\mf{g}_{\mathrm{nl},j}\psi_i)(\vr_\mu) 
\left(\sum_{c=1}^{N_c} 
\wt{\zeta}_{c \mu } (\vr) \prod_{c'\ne c} \frac{\varepsilon_i -
\wt{\varepsilon}_{c'}}{\wt{\varepsilon}_c -
\wt{\varepsilon}_{c'}} \right) \psi_{i}^{*}(\vr) +
  \mathrm{h.c.} \right].
  \end{split}
  \label{eqn:X0rg3-diag}
\end{equation}
The diagonal part of Eq.~\eqref{eqn:X0sg2} reads
\begin{equation}
\begin{split}
\left( \mf{X}_{0}^{(s)} \mf{g}_{j} \right)(\vr, \vr) \approx&
\sum_{\mu=1}^{N_{\mu}} W^{(s)}_{\mu} (\vr) (\Pi_\mu^{T} g_{\mathrm{loc},j}) 
\\
&+ \left[ \sum_{i=1}^{N_{\mathrm{cut}}} \sum_{\mu=1}^{N_{\mu}}
(\mf{g}_{\mathrm{nl},j} \psi_{i})(\vr_{\mu}) \left( \sum_{p=1}^{N_{p}}
\wt{\zeta}^{(s)}_{2,p\mu} (\vr) \omega_{p} (z_{p}-\varepsilon_{i})^{-1}
\right) \psi_{i}^{*} (\vr)  + \mathrm{h.c.} \right] \\
&+ \sum_{i=1}^{N_{\mathrm{cut}}} \sum_{\mu=1}^{N_{\mu}}
(\mf{g}_{\mathrm{nl},j} \psi_{i})(\vr_{\mu}) \left( \sum_{p=1}^{N_{p}}
\wt{\zeta}^{(s)}_{1,p\mu} (\vr) \omega_{p} (z_{p}-\varepsilon_{i})^{-1}
\right) \psi_{i}^{*} (\vr).
\end{split}
\label{eqn:X0sg2-diag}
\end{equation}

%\LL{TODO: Discuss how split representation helps to reduce the number of Dyson iterations.}

The reduced Dyson equation~\eqref{eqn:dysonX03} can be
readily solved using the same adaptive compression strategy in
section~\ref{subsec:acp}. More specifically, we can replace $\mf{g}_{j}$
by the local potential $\mathrm{diag}[f_{\mathrm{hxc}} u_{j}]$, and only
take the diagonal elements in Eq.~\eqref{eqn:X0rg3} and~\eqref{eqn:X0sg2}
to obtain $\chi_{0} f_{\mathrm{hxc}} u_{j}$. 
Moreover, since both the regular part $\chi_0^{(r)}$ and the singular part $\chi_0^{(s)}$ preserve a low-rank nature, Sherman-Morrison-Woodbury formula can still be used in the fixed point iteration. The separated treatment of the singular and regular parts reduces the error of the compressed $\chi_0$ as in Eq. \eqref{eqn:eta_chi0}. Therefore it also accelerates the convergence of the Dyson equation. The complete iteration process to solve the Dyson equations is defined in Alg.~\ref{alg:overall}.

% \begin{equation} \label{eqn:ShermanMorrison}
%                       \begin{split}
%                       U^{k+1} &= \left( I - W^{k}(\Pi^{k})^T f_\mathrm{hxc} \right)^{-1} U_0 \\
%                       &= U_0 +  W^{k} \left(I-(\Pi^{k})^T f_\mathrm{hxc} W^{k}\right)^{-1}(\Pi^{k})^T f_\mathrm{hxc} U_0
%                       \end{split}
%                       \end{equation}

\begin{algorithm}[ht]
        \small
        \DontPrintSemicolon
        \caption{Computing $U := [u_{j}]$ with the split representation of adaptively compressed polarizability operator.}
        \label{alg:overall}
        
        \KwIn{
                \begin{minipage}[t]{4in}  \end{minipage}\\
                 $\{\mf{g}_j\}$.  Stopping criterion $\delta$.
                
                Eigenpairs corresponding to occupied orbitals 
                $\{\psi_i,\varepsilon_i\}, i = 1,\ldots,\wt{N}_{\mathrm{cut}}$ 
        }
        \KwOut{
                \begin{minipage}[t]{4in}  
                        $U\approx \chi G$
                \end{minipage}
        } 
        \begin{enumerate}[leftmargin = *]
                \item Compute $U_0 := [u_{0,j}]$ using Eq.~\eqref{eqn:X0rg3} and~\eqref{eqn:X0sg2} (only the diagonal elements).
                \item \textbf{Do}  
                \begin{enumerate}
                        \item Replace $\{\mf{g}_j\}$ with $\mathrm{diag}[f_{\mathrm{hxc}} u^{k}_{j}]$ to obtain $W^{(r)k}$ and $W^{(s)k}$ and $\Pi^k$ in Eq.~\eqref{eqn:Wr} and~\eqref{eqn:Ws} . Define $W^{k} = W^{(s)k} + W^{(r)k}$. \\
                        \item Update $U^{k+1}$ using Sherman-Morrison-Woodbury formula
                        \begin{equation*}
                        \begin{split}
                        U^{k+1} &= \left( I - W^{k}(\Pi^{k})^T f_\mathrm{hxc} \right)^{-1} U_0 \\
                        &= U_0 +  W^{k} \left(I-(\Pi^{k})^T f_\mathrm{hxc} W^{k}\right)^{-1}(\Pi^{k})^T f_\mathrm{hxc} U_0
                        \end{split}
                        \end{equation*}.\\
                        \item $k\leftarrow k+1$
                \end{enumerate}
                \textbf{until} $\frac{\|U^{k} - U^{k-1}\|}{ \|U^{k-1} \|} < \delta$ or maximum number of iterations is reached.\\
        \end{enumerate}
        \label{alg:XgspACP}
\end{algorithm}

Once the self-consistent
$\mf{u}_{j}(\vr,\vr)$ are obtained, one can formally reconstruct
$\mf{u}(\vr,\vr')$ by using the split representation again in
Eq.~\eqref{eqn:X0rg3} and~\eqref{eqn:X0sg2}.  Finally $\mf{u}_{j}$ will be
integrated with $\mf{g}_{j'}$ as in Eq.~\eqref{eqn:SecDeriv} to compute
the Hessian matrix for phonon calculations, which will be further discussed in detail in the next section.

\subsection{Phonon Calculation}

For the purpose of phonon calculation, $\mf{u}_{j}$ (representing a component of $\frac{\partial P}{\partial \vR_I}$) will be
integrated with $\mf{g}_{j'}$ (representing a component of $\frac{\partial V_{\mathrm{ion}}}{\partial \vR_J} $) as in Eq.~\eqref{eqn:SecDeriv} to compute
the Hessian matrix for phonon calculations. The integration with local components can be readily computed once the self-consistent response $u_j(\vr)$ is obtained by solving the reduced Dyson equation. The integration with nonlocal components $\mf{g}_{\mathrm{nl},j}$ would require the construction of $\mf{u}(\vr,\vr')$. However since $\mf{g}_{\mathrm{nl},j}$ is compactly supported, one could avoid the full construction of $\mf{u}(\vr,\vr')$ by embedding the integration process into the construction of $\mf{u}(\vr,\vr')$. This is important for maintaining the reduced scaling of the algorithm. 

In this section, we show the construction of integral in Eq.~\eqref{eqn:chiterm}. For simplicity, the indexes $I, J$ are ignored. Starting from the Dyson equation,
\begin{equation}
        \mf{u}(\vr, \vr') = (\mf{X}_{0} \mf{g}) (\vr, \vr') +  (\mf{X}_{0}\mf{f}_{\mathrm{hxc}}\mf{u})(\vr,\vr'),
        \label{eqn:Reconstruction_Dyson}
\end{equation}
an element of the Hessian matrix requires calculation of 
\begin{equation}
        \int \mf{g}(\vr, \vr') \mf{u}(\vr, \vr') \ud \vr \ud \vr'  =\int  \left[ \mf{g}(\vr, \vr')  (\mf{X}_{0} \mf{g}) (\vr, \vr') + \mf{g}(\vr, \vr')  (\mf{X}_{0}\mf{f}_{\mathrm{hxc}}\mf{u})(\vr,\vr')  \right] \ud \vr \ud \vr'.
        \label{eqn:Reconstruction_1}
\end{equation}
Recall that $\mf{g}(\vr, \vr') = g_{\mathrm{loc}}(\vr) \delta (\vr' - \vr) + \mf{g}_\mathrm{nl} (\vr, \vr')$, the integral for the local part of can be easily calculated (letting $u(\vr)= \mf{u}(\vr, \vr)$)
\begin{equation}
\int \mf{g}_\mathrm{loc}(\vr, \vr') \mf{u}(\vr, \vr') \ud \vr \ud \vr'  = \int g_{\mathrm{loc}}(\vr) u(\vr) \ud \vr.  
\label{eqn:Reconstruction_2}
\end{equation}

For the non-local potential, using Eq.~\eqref{eqn:X0rg3} and Eq.~\eqref{eqn:X0sg2}, we have
\begin{equation}
     \int \mf{g}_\mathrm{nl}(\vr, \vr') \mf{u}(\vr, \vr') \ud \vr \ud \vr'  = \int \mf{g}_{\mathrm{nl}}(\vr, \vr') \left[(\mf{X}_{0} \mf{g}) (\vr, \vr') +  (\mf{X}_{0}\mf{f}_{\mathrm{hxc}}\mf{u})(\vr,\vr') \right] \ud \vr \ud \vr'.
     \label{eqn:nl-recon}
\end{equation}

Recall that $(\mf{f}_\mathrm{hxc} \mf{u}) (\vr, \vr') = \delta(\vr-\vr') \int f_\mathrm{hxc}(\vr, \vr'') u(\vr'') \ud \vr'' $. So $(\mf{f}_\mathrm{hxc} \mf{u}) (\vr, \vr')$ behaves as a local potential $g_\mathrm{loc}$ when applying $\mf{X}_{0}$ to it. So the integral in Eq.~\eqref{eqn:nl-recon} breaks down to four parts: 

\begin{equation}
\begin{split}
        &\int  \mf{g}_{\mathrm{nl}}(\vr, \vr')  (\mf{X}_{0}^{(r)} \mf{g}) (\vr, \vr') \ud \vr \ud \vr' \\ 
        =& \int -\sum_{l=1}^{L} \gamma_{l} (b_{l} (\vr) \ud b_{l}^* (\vr') + \ud b_{l} (\vr) b_{l}^* (\vr') ) (\mf{X}_{0}^{(r)} \mf{g} ) (\vr, \vr') \ud \vr \ud \vr'  \\
        = & -\int \ud \vr \ud \vr' \sum_{l=1}^{L} \gamma_{l} (b_{l} (\vr) \ud b_{l}^* (\vr') + \ud b_{l} (\vr) b_{l}^* (\vr') ) \sum_{\mu=1}^{N_\mu} \mf{W}_\mu^{(r)}[\mf{g}] (\vr, \vr') (\Pi_\mu[\mf{g}]^T g_{\mathrm{loc}}) \\
        & + \left[-\sum_{l=1}^{L} \gamma_{l} \sum_{i=1}^{N_{\mathrm{cut}}}\sum_{\mu=1}^{N_{\mu}} f_{i} 
        (\mf{g}_{\mathrm{nl}}\psi_i)(\vr_\mu) 
        \left(\sum_{c=1}^{N_c} 
        \int \ud \vr \wt{\zeta}_{c \mu }(\vr) b_{l}(\vr) \prod_{c'\ne c} \frac{\varepsilon_i -
                \wt{\varepsilon}_{c'}}{\wt{\varepsilon}_c -
                \wt{\varepsilon}_{c'}} \right) \int \ud \vr \psi_{i}^{*}(\vr') \ud b_{l}^* (\vr') \right. \\
        &\left. - \sum_{l=1}^{L} \gamma_{l} \sum_{i=1}^{N_{\mathrm{cut}}}\sum_{\mu=1}^{N_{\mu}} f_{i} 
        (\mf{g}_{\mathrm{nl}}\psi_i)(\vr_\mu) 
        \left(\sum_{c=1}^{N_c} 
        \int \ud \vr \wt{\zeta}_{c \mu }(\vr) \ud b_{l}(\vr) \prod_{c'\ne c} \frac{\varepsilon_i -
                \wt{\varepsilon}_{c'}}{\wt{\varepsilon}_c -
                \wt{\varepsilon}_{c'}} \right) \int \ud \vr \psi_{i}^{*}(\vr') b_{l}^* (\vr') \right] \\
        & + \text{h.c. of previous bracket}
\end{split}
\label{eqn:Reconstruction_r}
\end{equation}

\begin{equation}
\begin{split}
&\int  \mf{g}_{\mathrm{nl}}(\vr, \vr')  (\mf{X}_{0}^{(s)} \mf{g}) (\vr, \vr') \ud \vr \ud \vr' \\ 
=& \int -\sum_{l=1}^{L} \gamma_{l} (b_{l} (\vr) \ud b_{l}^* (\vr') + \ud b_{l} (\vr) b_{l}^* (\vr') ) (\mf{X}_{0}^{(s)} \mf{g} ) (\vr, \vr') \ud \vr \ud \vr'  \\
= & -\int \ud \vr \ud \vr' \sum_{l=1}^{L} \gamma_{l} (b_{l} (\vr) \ud b_{l}^* (\vr') + \ud b_{l} (\vr) b_{l}^* (\vr') ) \sum_{\mu=1}^{N_\mu} \mf{W}_{\mu}^{(s)} [\mf{g}](\vr, \vr') (\Pi_\mu[\mf{g}]^T g_{\mathrm{loc}}) \\
& + \left[ -\sum_{l=1}^{L} \gamma_{l}  \sum_{i=1}^{N_{\mathrm{cut}}} \sum_{\mu=1}^{N_{\mu}}
(\mf{g}_{\mathrm{nl},j} \psi_{i})(\vr_{\mu}) \left( \sum_{p=1}^{N_{p}}
\int \ud \vr \wt{\zeta}^{(s)}_{2,p\mu} (\vr) b_{l}(\vr) \omega_{p} (z_{p}-\varepsilon_{i})^{-1}
\right) \int \ud \vr'  \psi_{i}^{*} (\vr') \ud b_{l}^* (\vr')  \right.\\
&  \left. -\sum_{l=1}^{L} \gamma_{l}  \sum_{i=1}^{N_{\mathrm{cut}}} \sum_{\mu=1}^{N_{\mu}}
(\mf{g}_{\mathrm{nl},j} \psi_{i})(\vr_{\mu}) \left( \sum_{p=1}^{N_{p}}
\int \ud \vr \wt{\zeta}^{(s)}_{2,p\mu} (\vr) \ud b_{l}(\vr) \omega_{p} (z_{p}-\varepsilon_{i})^{-1}
\right) \int \ud \vr' \psi_{i}^{*} (\vr')b_{l}^*(\vr')  \right] \\
& + \text{h.c. of previous bracket} \\
& -\sum_{l=1}^{L} \gamma_{l}  \sum_{i=1}^{N_{\mathrm{cut}}} \sum_{\mu=1}^{N_{\mu}}
(\mf{g}_{\mathrm{nl},j} \psi_{i})(\vr_{\mu}) \left( \sum_{p=1}^{N_{p}}
\int \ud \vr \wt{\zeta}^{(s)}_{1,p\mu}(\vr) b_{l}(\vr) \omega_{p} (z_{p}-\varepsilon_{i})^{-1}
\right) \int \ud \vr' \psi_{i}^{*}(\vr') \ud b_{l}^*(\vr')\\
& -\sum_{l=1}^{L} \gamma_{l}  \sum_{i=1}^{N_{\mathrm{cut}}} \sum_{\mu=1}^{N_{\mu}}
(\mf{g}_{\mathrm{nl},j} \psi_{i})(\vr_{\mu}) \left( \sum_{p=1}^{N_{p}}
\int \ud \vr \wt{\zeta}^{(s)}_{1,p\mu}(\vr) \ud b_{l}(\vr) \omega_{p} (z_{p}-\varepsilon_{i})^{-1}
\right) \int \ud \vr' \psi_{i}^{*}(\vr') b_{l}^*(\vr') 
\end{split}
\label{eqn:Reconstruction_s}
\end{equation}

\begin{equation}
\begin{split}
        &\int  \mf{g}_{\mathrm{nl}}(\vr, \vr')  (\mf{X}_{0}^{(r)} \fhxc \mf{u}) (\vr, \vr') \ud \vr \ud \vr' \\ 
        =& \int -\sum_{l=1}^{L} \gamma_{l} (b_{l} (\vr) \ud b_{l}^* (\vr') + \ud b_{l} (\vr) b_{l}^* (\vr') ) (\mf{X}_{0}^{(r)} \fhxc \mf{u} ) (\vr, \vr') \ud \vr \ud \vr'  \\
        = & -\int \ud \vr \ud \vr' \sum_{l=1}^{L} \gamma_{l} (b_{l} (\vr) \ud b_{l}^* (\vr') + \ud b_{l} (\vr) b_{l}^* (\vr') ) \sum_{\mu=1}^{N_\mu} \mf{W}_\mu^{(r)} [\fhxc \mf{u}](\vr, \vr') (\Pi_\mu[\fhxc \mf{u}]^T f_\mathrm{hxc}u)
\end{split}
\label{eqn:Reconstruction_r2}
\end{equation}

\begin{equation}
\begin{split}
&\int  \mf{g}_{\mathrm{nl}}(\vr, \vr')  (\mf{X}_{0}^{(s)}  \fhxc \mf{u}) (\vr, \vr') \ud \vr \ud \vr' \\ 
=& \int -\sum_{l=1}^{L} \gamma_{l} (b_{l} (\vr) \ud b_{l}^* (\vr') + \ud b_{l} (\vr) b_{l}^* (\vr') ) (\mf{X}_{0}^{(s)} \fhxc \mf{u} ) (\vr, \vr') \ud \vr \ud \vr'  \\
= & -\int \ud \vr \ud \vr' \sum_{l=1}^{L} \gamma_{l} (b_{l} (\vr) \ud b_{l}^* (\vr') + \ud b_{l} (\vr) b_{l}^* (\vr') ) \sum_{\mu=1}^{N_\mu} \mf{W}_{\mu}^{(s)}[\fhxc \mf{u}] (\vr, \vr') (\Pi_\mu[\fhxc \mf{u}]^T  f_\mathrm{hxc}u) 
\end{split}
\label{eqn:Reconstruction_s2}
\end{equation}
We remark that the $\mf{W}$ quantity depends on the tensors to which $\mf{X}_0$ is applied.
Note that in Eqs.~\eqref{eqn:Reconstruction_r},
～\eqref{eqn:Reconstruction_s}, \eqref{eqn:Reconstruction_r2}, \eqref{eqn:Reconstruction_s2}, terms like $\int \ud \vr' \psi_{i}^{*}(\vr') b_{l}^*(\vr')$ appear many times, hence computing and storing them is necessary. Also one important fact is that $\mf{g}_{\mathrm{nl},j} \psi_{i}(\vr_{\mu})$ is only non-zero for several $\vr_{\mu}$. This would result in a ``fake'' summation of $N_\mu$, which is essential in reducing the complexity. Computation of Eq.~\eqref{eqn:Reconstruction_r} and Eq.~\eqref{eqn:Reconstruction_s} is only $\Or(N_e)$. The complexity is discussed in detail in the following section.

\subsection{Complexity}

In this section we analyze the complexity of phonon calculation using the split representation of ACP formulation, especially those related to nonlocal pseudopotential.

The first part of the algorithm is to compute the diagonal elements $u_{0,j}$ in Eq.~\eqref{eqn:X0rg3-diag} and~\eqref{eqn:X0sg2-diag}.  For the local pseudopotential, the cost of constructing $W^{(r)}$ and $W^{(s)}$ is $\Or(N_\mu N_{\mathrm{cut} }N_c N_g)\sim \Or(N_e^3)$ and $\Or(N_\mu N^{}_{\mathrm{cut} }N_p N_g)\sim \Or(N_e^3)$ respectively, since $N_{\mu},N_{\mathrm{cut}},N_g \sim \Or(N_e)$, and $N_c,N_p\sim \Or(1)$. Note that the construction of $W^{(r)},W^{(s)}$ does not depend on the index $j$, hence there is no factor of $dN_A$ involved. For the nonlocal pseudopotential, as is discussed in Section \ref{sec:CompRegular},  each nonlocal component of $\mf{g}_j$ is compactly supported in the real space. Denote $N_b$ as the grid points for the support of $\mf{g}_{\mathrm{nl},j}$. Hence for each $\mf{g}_{\mathrm{nl},j}$ there are only $N_b \sim \Or(1)$ number of points $\mathrm{r}_\mu$ that contributes to $(\mf{g}_{\mathrm{nl},j}
\psi_{i})(\vr_{\mu})$. So the cost associated with the nonlocal contribution is $\Or(dN_A N_{\mathrm{cut} } N_b N_c N_g)\sim \Or(N_e^3)$ in Eq.~\eqref{eqn:X0rg3} and $\Or( dN_A N_{\mathrm{cut}} N_b N_p N_g)\sim \Or(N_e^3)$ in Eq.~\eqref{eqn:X0sg2}. Note that the $dN_A$ factor comes from the fact that $\mf{g}_{\mathrm{nl},j} \psi_i (\vr_\mu)$ depends on index $j=1,2,\ldots,dN_A$.

In every iteration step when solving the reduced Dyson equation, the complexity of the construction of $W^k$ still cost $\Or(N_e^3)$, as we just replaced $\mf{g}_j$ by $\mathrm{diag}[f_{\mathrm{hxc}} u^{k}_{j}]$. Using Sherman-Morrison-Woodbury formula, the update of $U^{k+1}$ cost $\Or(N_g N_\mu d N_A + N_\mu^3 + N_\mu^2 dN_A)\sim \Or(N_e^3)$. In practice, we observe we observe that the number of iterations does not increase with respect to the system size.  To summarize, we know that the computation of $u_j(\vr)$ cost $\Or(N_e^3)$ in total.

In order to assemble the information stored in $\mf{u}_j$ to obtain the dynamical matrix for phonon calculations, $\mf{u}_{j}$ will be
integrated with $\mf{g}_{j'}$ as in Eq.~\eqref{eqn:SecDeriv}. Before we move on to further discussion, we note that $\mf{u}_j(\vr,\vr'), \mf{W}_\mu(\vr,\vr')$ are never constructed or stored. They are only stored in its factorized format. The integration with local components can be readily computed once the self-consistent response $u_j(\vr)$ is obtained by solving the reduced Dyson equation. The corresponding cost is  $\Or(d^2 N_A^2 N_g)$. The integration with nonlocal components $\mf{g}_{\mathrm{nl},j}$ would require certain off-diagonal entries $\mf{u}(\vr,\vr')$. However since $\mf{g}_{\mathrm{nl},j}$ is compactly supported, one could avoid the full construction of $\mf{u}(\vr,\vr')$ by embedding the integration process into the construction of $\mf{u}(\vr,\vr')$. As shown in Eqs.~\eqref{eqn:Reconstruction_r} and \eqref{eqn:Reconstruction_s}, the complexity for this integration is $\Or(d^2 N_a^2 N_b N_{\mathrm{cut}} N_c  + 2 d^2 N_a^2 N_\mathrm{cut} N_b N_c) \sim \Or(N_e^3)$ and $\Or(d^2 N_a^2 N_b^2 N_{\mathrm{cut}} N_p  + 2 d^2 N_a^2 N_\mathrm{cut} N_b N_p) \sim \Or(N_e^3)$, respectively. As for Eqs.~\eqref{eqn:Reconstruction_r2} and \eqref{eqn:Reconstruction_s2}, the complexity is $\Or(d^2 N_a^2 N_b N_{\mathrm{cut}} N_c) \sim \Or(N_e^3)$ and $\Or(d^2 N_a^2 N_b N_{\mathrm{cut}} N_p) \sim \Or(N_e^3)$, respectively. Diagonalizing the Hessian matrix costs $\Or(N_a^3)$. In summary, the complexity of phonon calculation scales as $\Or(N_e^3)$. This is further confirmed by numerical examples in 1D in the following section. Table~\ref{tab:Complexity} summarizes the complexity of all computation steps of split ACP.

\begin{table}[ht]
        \begin{center}
                \begin{tabular}{c|c|c} \hline
                        Step & Equation & Complexity  \\ \hline 
           Interpolation decomposition & Eq.~\eqref{eqn:X0interpdecompose}  & \tabincell{c}{$\Or(N_g d N_A N_\mathrm{cut})$ \\ $ + \Or( N_g N_{\mathrm{cut}} N_\mu)$ } \\ \hline
           \tabincell{c}{Diagonal element construction \\ regular part} & Eq.~\eqref{eqn:X0rg3-diag} &  \tabincell{c}{$\Or(N_\mu N_{\mathrm{cut} }N_c N_g)$ \\  $+ \Or(dN_A N_{\mathrm{cut}} N_b N_c N_g)$ } \\ \hline
           \tabincell{c}{Diagonal element construction \\ Singular part } & Eq.~\eqref{eqn:X0sg2-diag} & \tabincell{c}{$\Or(N_\mu N^{}_{\mathrm{cut} }N_p N_g)$ \\ $ + \Or( dN_A N^{}_{\mathrm{cut} }N_b N_p N_g)$ } \\ \hline
           The Dyson equation update & \tabincell{c}{Step 2.(b) \\ in Alg.~\ref{alg:XgspACP} }  & \tabincell{c}{$\Or(N_g N_\mu d N_A)$ \\ $ + \Or(N_\mu^3 + N_\mu^2 dN_A)$  } \\ \hline
            \tabincell{c}{Reconstruction \\ local potential} & Eq.~\eqref{eqn:Reconstruction_2} & \tabincell{c}{$\Or(d^2 N_A^2 N_g)$} \\ \hline
           \tabincell{c}{Reconstruction \\ nonlocal pseudopotential} &
           \tabincell{c}{Eq.~\eqref{eqn:Reconstruction_r} \\
           Eq.~\eqref{eqn:Reconstruction_s} \\
           Eq.~\eqref{eqn:Reconstruction_r2} \\ Eq.~\eqref{eqn:Reconstruction_s2} }& \tabincell{c}{
           $\Or(d^2 N_a^2 N_b N_{\mathrm{cut}} N_c  + 2 d^2 N_a^2 N_\mathrm{cut} N_b N_c) $ \\
           $\Or(d^2 N_a^2 N_b N_{\mathrm{cut}} N_p  + 2 d^2 N_a^2 N_\mathrm{cut} N_b N_p) $ \\
           $\Or(d^2 N_a^2 N_b N_{\mathrm{cut}} N_c)$ \\
           $\Or(d^2 N_a^2 N_b N_{\mathrm{cut}} N_p)$
           \\
           } \\ \hline
                \end{tabular}
        \end{center}
        \caption{ Summary of the complexity of each component of the split ACP algorithm.}
        \label{tab:Complexity}
\end{table}

\section{Numerical examples}\label{sec:numer}

In this section, we demonstrate the performance of split ACP 
and compare it with DFPT and finite difference (FD) through two examples. 
The first example consists of a 1D reduced Hartree-Fock model
problem that can be tuned to resemble a metallic system. 
%We use a uniform grid and the second-order finite difference scheme for spacial discretization. Period boundary condition is adopted. 
The second one is a 3D aluminum cluster calculation 
performed using KSSOLV~\cite{YangMezaLeeEtAl2009}, which is a MATLAB toolbox for solving Kohn-Sham equations for small molecules and solids in three-dimensions. KSSOLV uses plane wave expansion to discretize the Kohn-Sham
equations. 
All calculations are carried out using the Berkeley Research Computing (BRC) High Performance Computing service. 
Each node consists of two Intel Xeon 10-core Ivy Bridge processors (20 cores
per node) and 64 GB of memory.

\subsection{1D reduced Hartree-Fock model with nonlocal pseudopotential}

The 1D reduced Hartree-Fock model was introduced by Solovej
\cite{Solovej1991}, and has been used for analyzing defects in solids
in e.g. \cite{CancesDeleurenceLewin2008,CancesDeleurenceLewin2008a}. The
simplified 1D model neglects the contribution of the
exchange-correlation term. As discussed in previous sections, the
presence of exchange-correlation functionals at LDA/GGA level does not
lead to essential difficulties in phonon calculations. 
Furthermore, the nonlocal pseudopotential in the 
Kleinman-Bylander form~\cite{KleinmanBylander:82} is added to this reduced model 
to test the availability for the split ACP to handle the case 
in presence of nonlocal potential.

The Hamiltonian in our 1D reduced Hartree-Fock model is given by 
\begin{equation}
        H [ \rho ] = -\frac{1}{2} \frac{d^2}{dx^2} 
    + \left[\int K(x,y) \left( \rho(y) + m(y) \right) \ud y\right] \delta(x,x')
    + \gamma \sum_I b(x-R_I)b^*(x'-R_I). 
        \label{eqn:1DHamil}
\end{equation}
Here $m(x) = \sum_{I} m_I(x - R_I)$ is the summation of pseudocharges. Each function $m_I(x)$ takes the form of a one-dimensional Gaussian 
\begin{align}
m_I(x) = - \frac{Z_I}{\sqrt{2\pi\sigma_I^2}} \exp \left( -\frac{x^2}{2\sigma_I^2}  \right),
\end{align}
where $Z_I$ is an integer representing the charge of the $I$-th nucleus. In our numerical simulation, we choose all $\sigma_I$ to be the same.
 
Instead of using a bare Coulomb interaction which diverges in 1D when $x$ is large, we use a Yukawa kernel as the regularized Coulomb kernel
\begin{equation}
K(x, y) = \frac{2\pi e^{-\kappa | x-y | }}{\kappa \epsilon_0}, \label{eqn:yukawa}
\end{equation}
which satisfies the equation
\begin{align}
-\frac{d^2}{dx^2}K(x,y) + \kappa^2 K(x,y) = \frac{4\pi}{\epsilon_0}\delta(x-y).
\end{align}
As $\kappa \rightarrow 0$, the Yukawa kernel approaches the bare Coulomb interaction given by the Poisson equation. The parameter $\epsilon_0$ is used so that the magnitude of the electron static contribution is comparable to that of the kinetic energy.
The ion-ion repulsion energy $E_{\mathrm{II}}$ is also computed using the Yukawa interaction $K$ in the model systems.

The last term in $H[\rho]$ represents the kernel of the nonlocal pseudopotential, 
which is the summation of rank-1 real symmetric operator with real valued function 
\begin{equation}
b(x) = \frac{1}{\sqrt{2\pi\sigma_b^2}} \exp \left( -\frac{x^2}{2\sigma_b^2}  \right). 
\end{equation}
$\gamma$ is a scaling factor used to control the magnitude of the nonlocal 
pseudopotential, which is, in practice, much smaller than the local pseudopotential.  

The parameters used in this model are chosen as follows. 
Atomic units are used throughout the discussion unless otherwise mentioned. 
For all systems tested in this subsection, the distance between each atom and its 
nearest neighbor is set to 2.4 a.u.
The Yukawa parameter $\kappa = 0.1$. 
The nuclear charge $Z_I$ is set to 1 for all atoms, 
and $\sigma_I$ is set to 0.3. 
The parameter $\epsilon_0$ is chosen to be  80 so that 
the reduced Hartree-Fock model can be tuned to resemble a metallic system. 
In the nonlocal pseudopotential, the scaling factor $\gamma = -0.01$, as well as 
$\sigma_b$ set to be 0.1 (this will cause the total energy to change by $1.47\%$). 
The temperature $T$ is set to be 5000 K to emphasize the influence of 
partial occupation. 
The Hamiltonian operator is represented in a plane wave basis set. 

\begin{figure}[ht]
        \begin{minipage}[h]{\linewidth}

                                \centering
                                \includegraphics[width=0.75\linewidth]{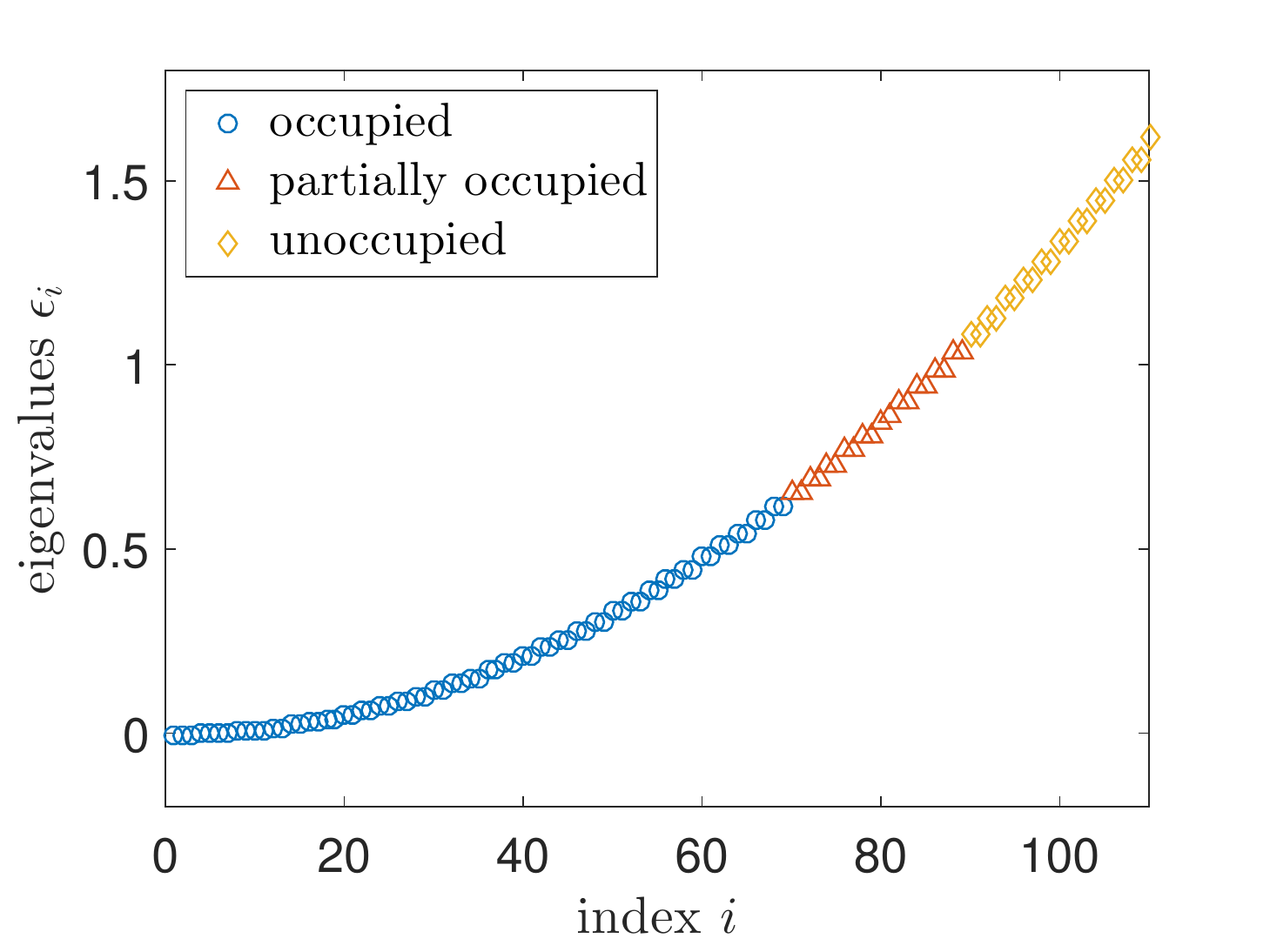}
                                \subcaption{eigenvalues}
                                \label{fig:1Dground:eigs}

        \end{minipage}
  \caption{Eigenvalues of the 1D system with $N_A = 80$. }        \label{fig:1Dground}
\end{figure}

For the system of size $N_A = 80$, the 110 smallest eigenvalues
are shown in Fig.~\ref{fig:1Dground}, and the corresponding occupational status near the chemical potential is shown in Fig.~\ref{fig:1Dground_cutoff}. 
There is no evident energy gap within 
the spectrum of the Hamiltonian. 
Orbitals can be partially occupied due to the finite temperature. 
Specifically, we identify an orbital to be (fully) occupied if the occupation number 
$f_i > 1-10^{-6}$, unoccupied if $f_i < 10^{-6}$, otherwise partially occupied. 
In this case, there are 20 partially occupied orbitals, whose eigenvalues are around 
the chemical potential. 
The total number of (fully) occupied and partially occupied orbitals $N_{\text{occ}}$ 
is 89, and we choose $N_{\mathrm{cut}} = N_{\text{occ}}$ for all the split ACP 
computations. Also we fix the number of pole expansion nodes $N_p$ to be 40 
unless otherwise mentioned.

In the ground state calculation, we use Anderson mixing~\cite{Anderson1965} for accelerating the self-consistent field  (SCF) iterations, 
and the linearized eigenvalue problems are solved by using the 
locally optimal block preconditioned conjugate gradient (LOBPCG) 
solver~\cite{Knyazev2001}. 
In DFPT, we use MINRES~\cite{PaigeSaunders1975} to solve the Sternheimer equations iteratively. 
The initial guess vectors for the solutions are obtained from previous iterations 
in the Dyson equation to reduce the number of matrix-vector multiplications. 
The same strategy for choosing the initial guess is implemented for the 
split ACP formulation as well. 
Anderson mixing is used to accelerate the convergence of Dyson equations in DFPT, 
and in split ACP we use the fixed point iteration with Sherman-Morrison-Woodbury 
formula. 

All numerical results of the split ACP method and FD approach below are benchmarked 
with results obtained from DFPT. We test the accuracy of the split 
ACP method in three different level: the diagonal elements
$\text{diag}(\mf{X}_{0}\mf{g})$, 
the diagonal elements of solution to Dyson equations $\text{diag}(\mf{X}\mf{g})$, and the phonon frequencies $\{\omega_k\}$. 
For the diagonal elements $\text{diag}(\mf{X}_{0}\mf{g})$ and $\text{diag}(\mf{X}\mf{g})$, we directly measure the relative $L^2$ error, 
defined as $\|\text{diag}(\mf{X}_{0}\mf{g})-\text{diag}(\wt{\mf{X}}_{0}\mf{g})\|_2/\|\text{diag}(\mf{X}_{0}\mf{g})\|_2$.
For the phonon frequencies, due to the presence of acoustic phonon modes for which 
$\omega_k$ is close to 0, instead of the relative error, 
we measure the absolute $L^{\infty}$ error defined as 
$\max_k|\omega_k - \widetilde{\omega}_k|$, where $\widetilde{\omega}_k$ is obtained 
from FD or split ACP. 
We also demonstrate the efficiency of the split ACP method by comparing the 
computational time and scaling of split ACP with that of DFPT and FD.

\begin{table}[ht]
        \begin{center}
                \begin{tabular}{c|c|c|c|c|c|c} \hline
                        \diagbox{$N_c$}{$N_\mu$} & $3N_{\text{occ}}$ & $4N_{\text{occ}}$ & $5N_{\text{occ}}$ & $6N_{\text{occ}}$ & $7N_{\text{occ}}$ & $8N_{\text{occ}}$\\ \hline
                        3  & 2.38E-02 & 2.17E-02 & 2.13E-02 & 2.12E-02 & 2.12E-02 & 2.12E-02 \\ \hline
            4  & 2.06E-02 & 9.43E-03 & 6.25E-03 & 6.21E-03 & 6.21E-03 & 6.21E-03 \\ \hline
            5  & 2.01E-02 & 7.88E-03 & 2.86E-03 & 2.85E-03 & 2.84E-03 & 2.84E-03 \\ \hline
            6  & 1.64E-02 & 6.76E-03 & 1.73E-03 & 1.65E-03 & 1.65E-03 & 1.65E-03 \\ \hline
            7  & 1.65E-02 & 9.30E-03 & 8.10E-04 & 6.85E-04 & 6.87E-04 & 6.87E-04 \\ \hline
            8  & 1.62E-02 & 9.07E-03 & 5.86E-04 & 2.53E-04 & 2.50E-04 & 2.50E-04 \\ \hline
            9  & 1.81E-02 & 7.24E-03 & 7.86E-04 & 1.51E-04 & 1.47E-04 & 1.47E-04 \\ \hline
            10 & 1.49E-02 & 6.53E-03 & 5.83E-04 & 7.99E-05 & 7.24E-05 & 7.24E-05 \\ \hline
                \end{tabular}
        \end{center}
        \caption{\REV{The relative $L^{2}$ error $\|\text{diag}(\mf{X}_{0}\mf{g})-\text{diag}(\wt{\mf{X}}_{0}\mf{g})\|_2/\|\text{diag}(\mf{X}_{0}\mf{g})\|_2$ for $\widetilde{N}_{\text{cut}}/N_{\text{cut}} \approx 1.06$ 
  with the effective gap $\widetilde{\varepsilon}_g / |\mc{I}| \approx 0.1408$.}
  }
        \label{tab:1D_accu_response1}
\end{table}
\begin{table}[ht]
        \begin{center}
                \begin{tabular}{c|c|c|c|c|c|c} \hline
                        \diagbox{$N_c$}{$N_\mu$} & $3N_{\text{occ}}$ & $4N_{\text{occ}}$ & $5N_{\text{occ}}$ & $6N_{\text{occ}}$ & $7N_{\text{occ}}$ & $8N_{\text{occ}}$\\ \hline
                        3  & 1.56E-02 & 8.52E-03 & 9.45E-04 & 7.42E-04 & 7.39E-04 & 7.39E-04 \\ \hline
            4  & 1.72E-02 & 7.79E-03 & 6.82E-04 & 1.02E-04 & 9.67E-05 & 9.67E-05 \\ \hline
            5  & 1.74E-02 & 9.49E-03 & 8.90E-04 & 6.00E-05 & 2.50E-05 & 2.50E-05 \\ \hline
            6  & 1.56E-02 & 7.80E-03 & 5.89E-04 & 7.06E-05 & 5.40E-06 & 5.38E-06 \\ \hline
            7  & 1.62E-02 & 9.07E-03 & 6.11E-04 & 5.51E-05 & 8.45E-07 & 8.42E-07 \\ \hline
            8  & 1.61E-02 & 9.04E-03 & 5.97E-04 & 4.73E-05 & 5.55E-07 & 3.21E-07 \\ \hline
            9  & 1.85E-02 & 9.08E-03 & 6.45E-04 & 4.52E-05 & 4.88E-07 & 3.20E-07 \\ \hline
            10 & 1.55E-02 & 9.52E-03 & 8.12E-04 & 5.72E-05 & 4.97E-07 & 3.20E-07 \\ \hline
                \end{tabular}
        \end{center}
        \caption{\REV{The relative $L^{2}$ error $\|\text{diag}(\mf{X}_{0}\mf{g})-\text{diag}(\wt{\mf{X}}_{0}\mf{g})\|_2/\|\text{diag}(\mf{X}_{0}\mf{g})\|_2$ for $\widetilde{N}_{\text{cut}}/N_{\text{cut}} \approx 1.28$ 
  with the effective gap $\widetilde{\varepsilon}_g / |\mc{I}| \approx 0.6777$.}
  }
        \label{tab:1D_accu_response2}
\end{table}

In Table~\ref{tab:1D_accu_response1} and~\ref{tab:1D_accu_response2}, we calibrate the 
accuracy of the split compression with different choices of 
the numbers of Chebyshev nodes $N_c$ and the numbers of columns $N_{\mu}$, 
for two different choices of $\wt{N}_{\mathrm{cut}}$, respectively. 
We measure the accuracy by relative $L^2$ error 
$\|\text{diag}(\mf{X}_{0}\mf{g})-\text{diag}(\wt{\mf{X}}_{0}\mf{g})\|_2/\|\text{diag}(\mf{X}_{0}\mf{g})\|_2$, 
and choose $N_{\mu} = lN_{\text{occ}}$ where $l = 3, 4, \cdots, 8$. 
Table~\ref{tab:1D_accu_response1} and~\ref{tab:1D_accu_response2} both show that, 
with a fixed number of Chebyshev nodes $N_c$, the error decreases monotonically with respect to $N_{\mu}$, until limited by the accuracy of the 
Chebyshev interpolation procedure. Similarly, with a fixed number of selected columns, 
the numerical accuracy improves as more Chebyshev nodes are used in interpolation 
until limited by the choice of $N_{\mu}$. 
Comparing Table~\ref{tab:1D_accu_response2} with Table~\ref{tab:1D_accu_response1}, 
we also find that numerical accuracy can be better with a larger 
$\widetilde{N}_{\mathrm{cut}}$. This is due to the increase of the effective energy gap $\widetilde{\varepsilon}_g$, 
which leads to a smaller numerical error in the Chebyshev interpolation procedure. 
For $\widetilde{N}_{\text{cut}}/N_{\text{cut}} \approx 1.28$, 
the relative $L^2$ error of $\chi_0G$ can be less than $10^{-6}$ for large enough 
$N_c$ and $N_{\mu}$. 

We further study how different choices of $N_c$ and $\widetilde{N}_{\text{cut}}$ 
affect the computational accuracy on $\mathrm{diag}(\mf{X}_0 \mf{g} )$. 
Here for all $N_c$ and $\widetilde{N}_{\text{cut}}$, $N_{\mu}$ is fixed to be 
$480 \approx 5.4 N_{\text{occ}}$ or $560 \approx 6.3 N_{\text{occ}}$. 
This is determined the same way as 
that in the regular ACP formulation so that 
$|\widetilde{R}_{N_{\mu}+1,N_{\mu}+1}| < \epsilon |\widetilde{R}_{1,1}| \leq |\widetilde{R}_{N_{\mu},N_{\mu}}|$ 
in Algorithm 2 in~\cite{LinXuYing2017}, with $\epsilon = 10^{-4}$ and $10^{-5}$, 
respectively. 
\begin{figure}[ht]
        \begin{minipage}[h]{\linewidth}
                        \begin{minipage}[t]{0.5\linewidth}
                                \centering
                                \includegraphics[width=0.98\linewidth]{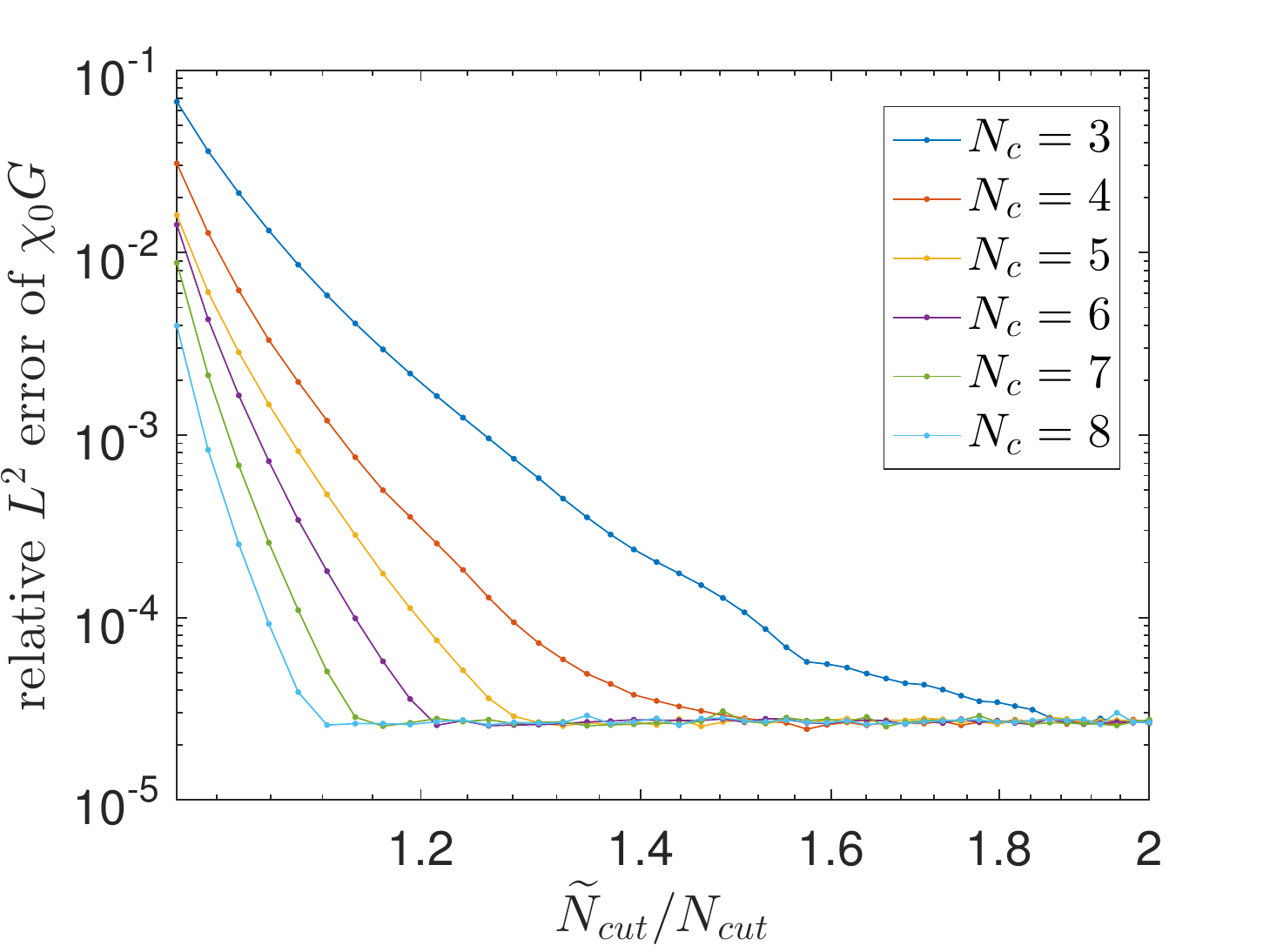}
                                \subcaption{$N_{\mu} \approx 5.4 N_{\text{occ}}$}
                                \label{fig:1DNcNtot_tau1e-4}
                        \end{minipage}%
                        \begin{minipage}[t]{0.5\linewidth}
                                \centering
                                \includegraphics[width=0.98\linewidth]{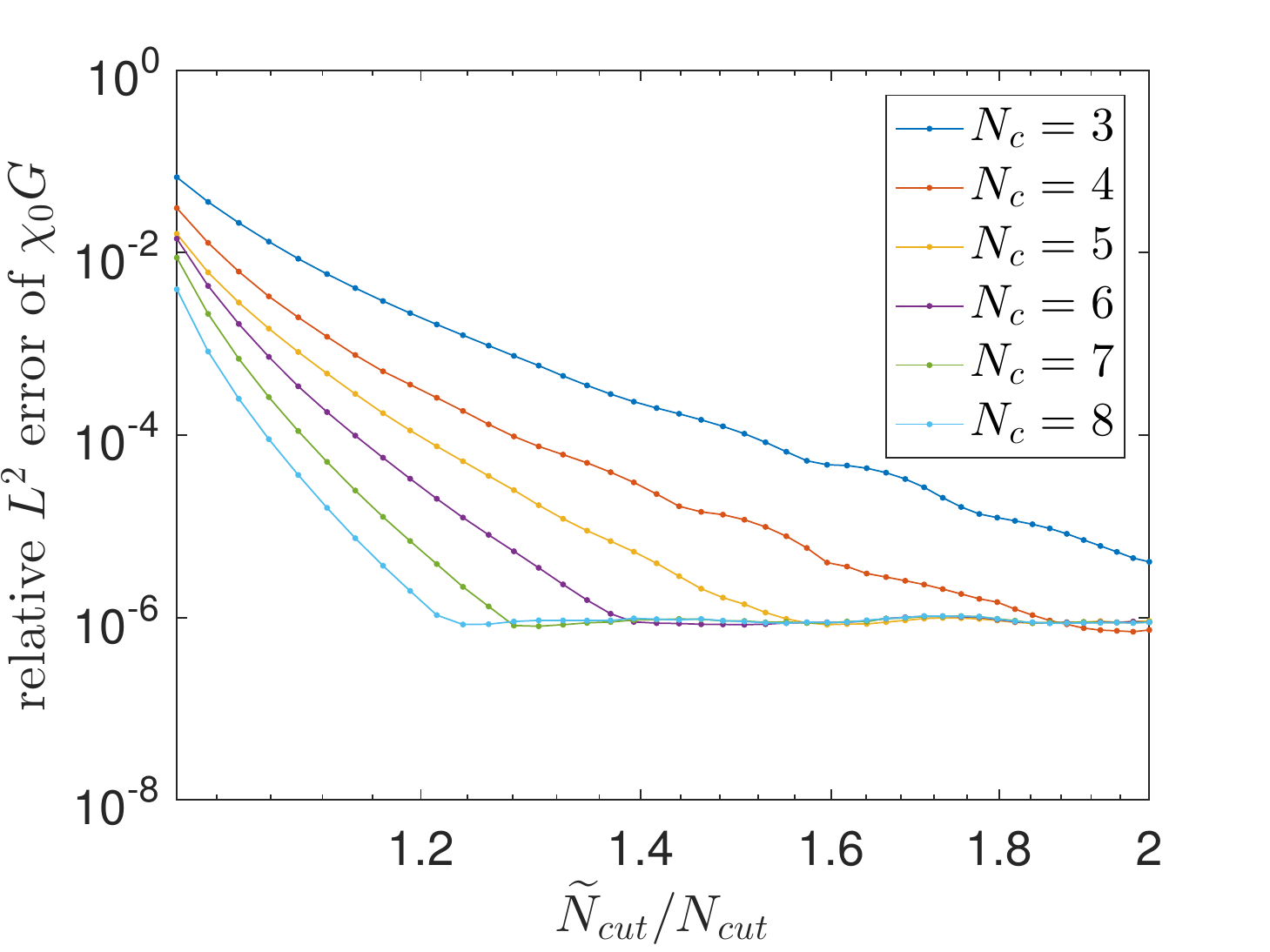}
                                \subcaption{$N_{\mu} \approx 6.3 N_{\text{occ}}$}
                                \label{fig:1DNcNtot_tau1e-5}
                        \end{minipage}
        \end{minipage}
  \caption{The relative $L^2$ errors 
  $\|\text{diag}(\mf{X}_{0}\mf{g})-\text{diag}(\wt{\mf{X}}_{0}\mf{g})\|_2/\|\text{diag}(\mf{X}_{0}\mf{g})\|_2$ under different
  $\widetilde{N}_{\text{cut}}$ and $N_c$}
        \label{fig:1DNcNtot}
\end{figure}
Fig.~\ref{fig:1DNcNtot} compares the relative $L^2$ errors 
$\|\text{diag}(\mf{X}_{0}\mf{g})-\text{diag}(\wt{\mf{X}}_{0}\mf{g})\|_2/\|\text{diag}(\mf{X}_{0}\mf{g})\|_2$ under different 
$\widetilde{N}_{\text{cut}}$ and $N_c$. We find that it can be sufficient to choose 
$\widetilde{N}_{\text{cut}} \leq 2N_{\text{cut}}$ to achieve the best accuracy possible
where further improvement is hindered by the the choice of $N_{\mu}$ 
(around $3\times 10^{-5}$ for $N_{\mu} \approx 5.4 N_{\text{occ}}$ 
and $1\times 10^{-6}$ for $N_{\mu} \approx 6.3 N_{\text{occ}}$).
Under the split ACP formulation, the number of Chebyshev nodes is 
significantly reduced. Specifically, 4-8 nodes can already perform fairly 
accurate calculation while no less than 20 nodes are needed in the regular 
ACP formulation. Furthermore, the more Chebyshev nodes are used, 
the smaller $\widetilde{N}_{\text{cut}}$ we can choose to achieve the same accuracy. 
For example, if 5 nodes are adopted in Chebyshev interpolation, 
we need to choose $\widetilde{N}_{\text{cut}}$ as large as $1.55 N_{\text{cut}}$ 
to achieve the best accuracy, while 
$\widetilde{N}_{\text{cut}} \approx 1.2 N_{\text{cut}}$ is sufficient if $N_c$ 
increases to 8. 

\begin{figure}[ht]
        \begin{center}
                \includegraphics[width=0.7\textwidth]{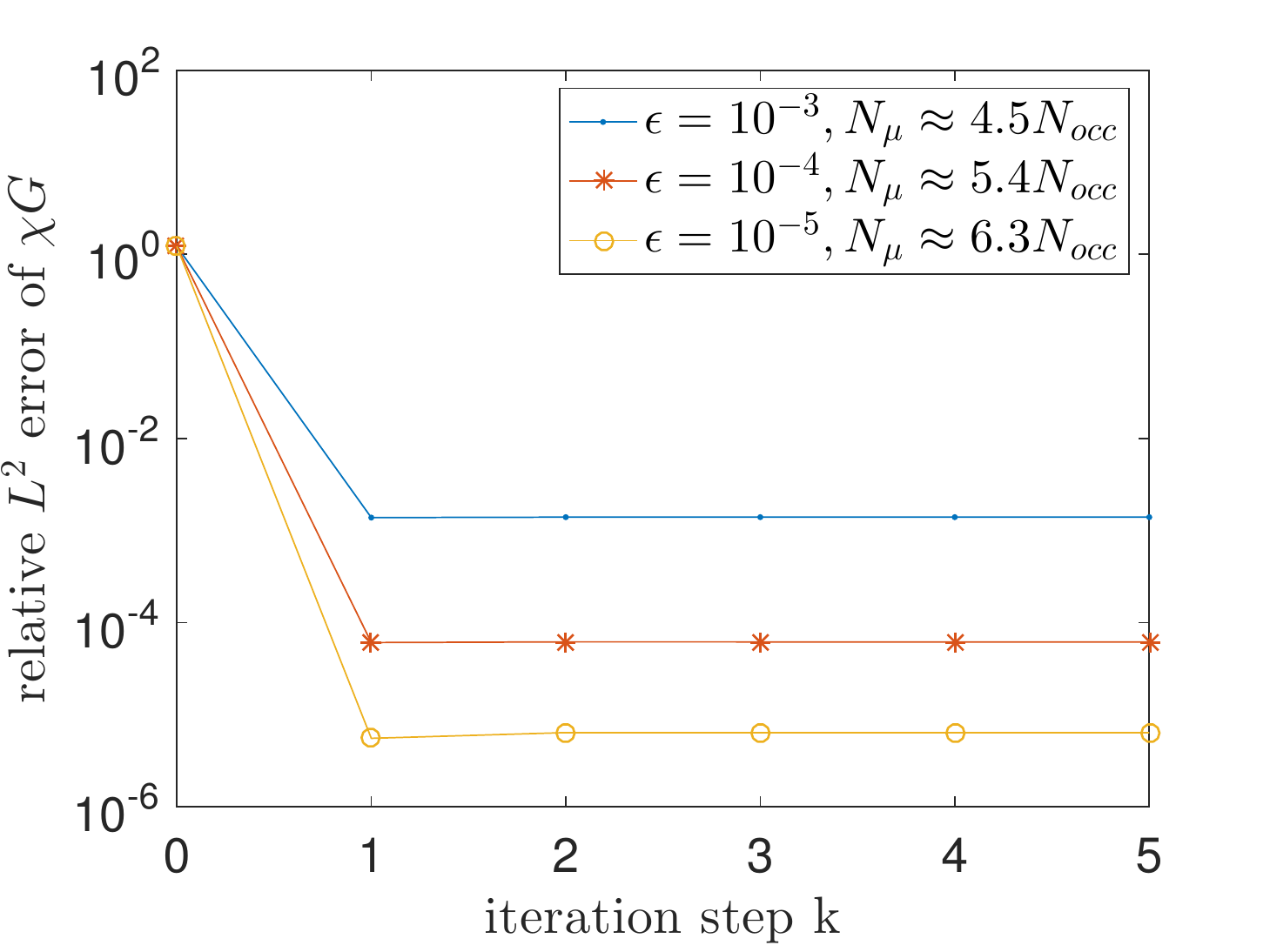}
        \end{center}
        \caption{Convergence for solving the Dyson equation using the split ACP formulation.}
        \label{fig:1DDyson}
\end{figure}
In order to demonstrate the effectiveness of the split representation, 
the relative $L^2$ error $\|\text{diag}(\mf{X}\mf{g})-\text{diag}(\wt{\mf{X}}\mf{g})\|_2/\|\text{diag}(\mf{X}\mf{g})\|_2$   during the fixed 
point iteration when solving Dyson equation is shown in Fig.~\ref{fig:1DDyson}. 
For each choice of $N_{\mu}$, numerical results show significant improvement after 
only one iteration, and the self-consistent iteration converges within two steps. 
After convergence, the error is around $1.4\times 10^{-3}$ for $\epsilon = 10^{-3}$, 
$6.2\times 10^{-5}$ for $\epsilon = 10^{-4}$, and 
$6.4\times 10^{-6}$ for $\epsilon = 10^{-5}$.

\begin{table}[ht]
        \begin{center}
                \begin{tabular}{c|c} \hline
                        Method and parameters  & $L^{\infty}$-norm error \\ \hline
            FD, $\delta = 0.01$   &  7.79E-05 \\ \hline
            split ACP, $N_p = 20$, $N_{\mu} \approx 5.4 N_{\text{occ}}$ for $\epsilon = 10^{-4}$ & 5.90E-05 \\ \hline
            split ACP, $N_p = 40$, $N_{\mu} \approx 6.3 N_{\text{occ}}$ for $\epsilon = 10^{-5}$ & 1.51E-05 \\ \hline
                \end{tabular}
        \end{center}
        \caption{ $L^{\infty}$ error of the phonon frequencies. 
    System size is $N_A = 80$. Chebyshev nodes $N_c = 5$ in split ACP. }
        \label{tab:1D_accu_phonon}
\end{table}

Next we compare the split ACP with DFPT and FD in terms of the accuracy of phonon frequencies.
Table~\ref{tab:1D_accu_phonon} presents $L^{\infty}$ error of the phonon spectrum 
obtained by FD and split ACP with different parameters benchmarked with that from DFPT. 
In the FD approach, the convergence tolerance for LOBPCG is set to be $10^{-8}$, 
and the SCF convergence tolerance is $10^{-10}$. 
$\delta$ denotes the perturbation of each atom position to the origin. 
We remark that further smaller $\delta$ can lead to slightly larger numerical error 
due to the numerical instability of FD approach, 
and the numerical error of FD approach is usually around $10^{-4}$. 
As for the split ACP, the same parameters for LOBPCG and SCF are chosen to 
converge the ground state calculation, and 5 nodes are used in the 
Chebyshev interpolation procedure. 
We find that it is sufficient to choose $N_p = 20$ and $N_{\mu} \approx 5.4 N_{\text{occ}}$ 
to achieve comparable accuracy with FD approach. 
Furthermore, with more nodes in pole expansion and more selected columns, 
the $L^{\infty}$ error of split ACP can be as small as around $10^{-5}$, 
in which case split ACP can be more accurate than FD approach. 

\begin{figure}[ht]
        \begin{minipage}[h]{\linewidth}
                        \begin{minipage}[t]{0.5\linewidth}
                                \centering
                                \includegraphics[width=0.98\linewidth]{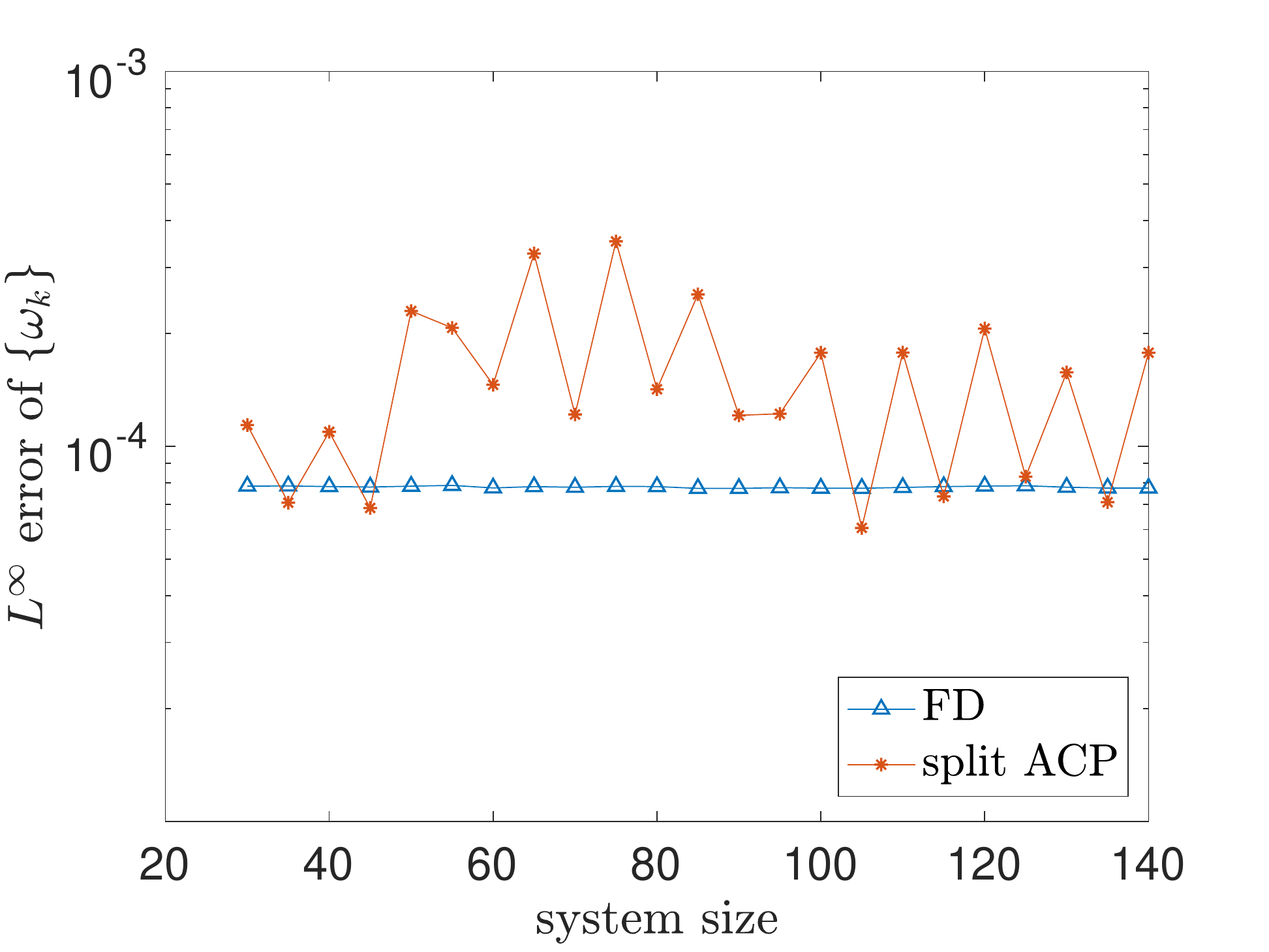}
                                \subcaption{}
                                \label{fig:1D_phonon:accu}
                        \end{minipage}%
                        \begin{minipage}[t]{0.5\linewidth}
                                \centering
                                \includegraphics[width=0.98\linewidth]{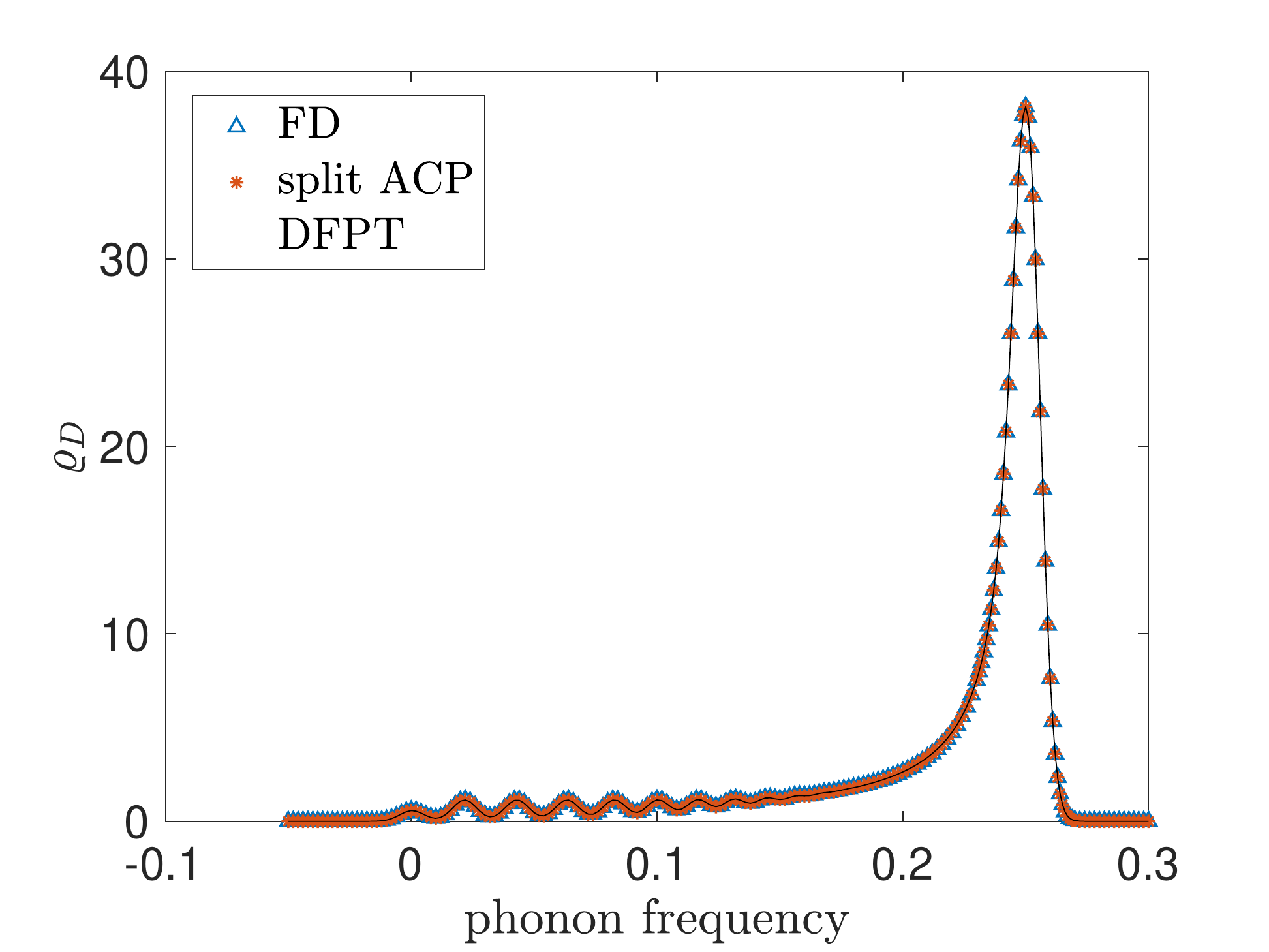}
                                \subcaption{}
                                \label{fig:1Dphonon:spec}
                        \end{minipage}
        \end{minipage}
  \caption{(a) $L^{\infty}$ error of the phonon frequencies $\{\omega_k\}$. 
  (b) Phonon spectrum for the 1D system.}
        \label{fig:1D_phonon}
\end{figure}

\begin{figure}[ht]
        \begin{minipage}[h]{\linewidth}
                \centering
                \includegraphics[width=0.7\linewidth]{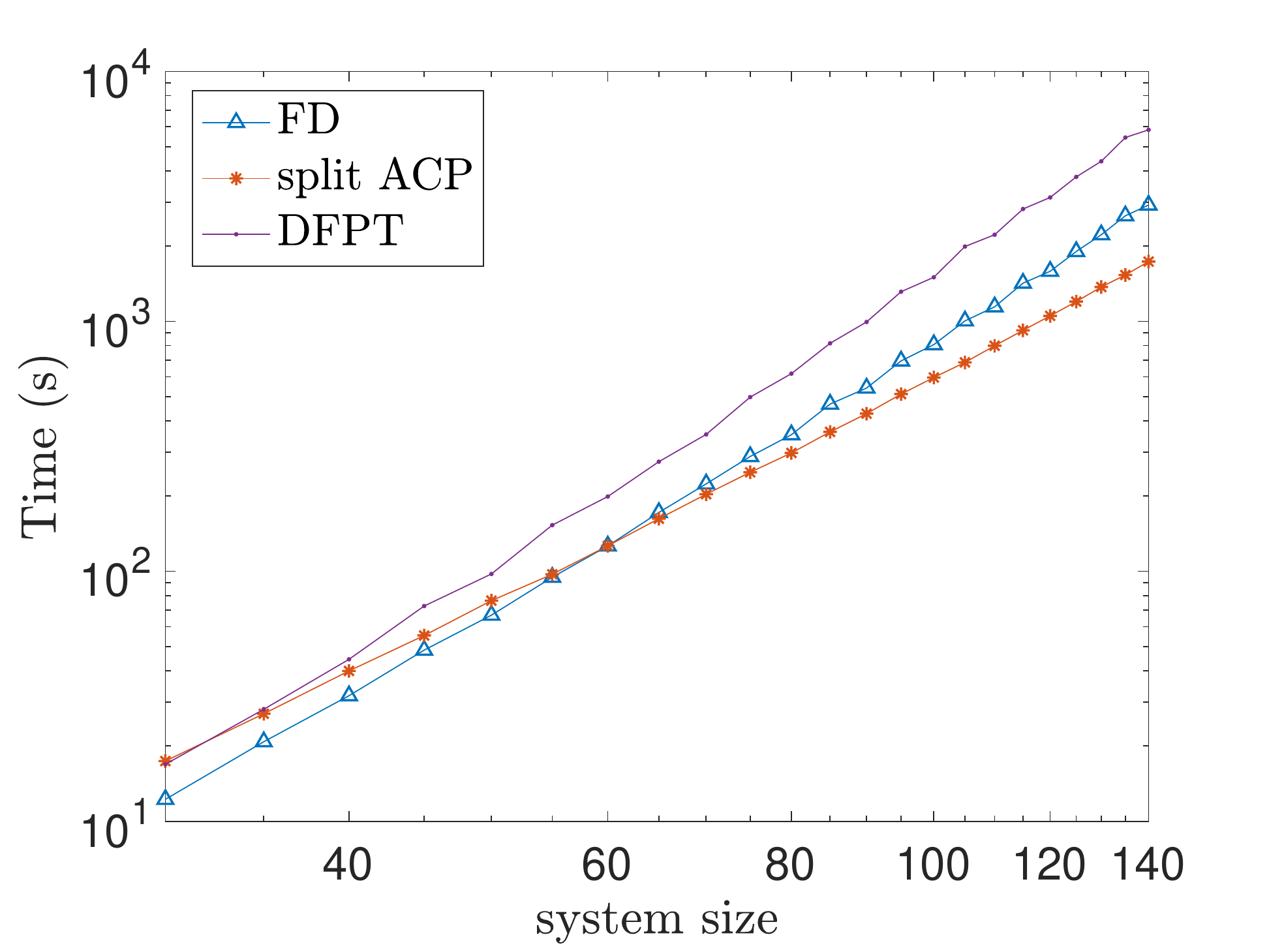}
        \end{minipage}
  \caption{Computational time of 1D examples. }
        \label{fig:1D_time}
\end{figure}

\begin{table}[ht]
        \begin{center}
                \begin{tabular}{c|c} \hline
                        Method  & Computational scaling \\ \hline
            DFPT  & 4.0036 \\ \hline
            FD    & 3.8057 \\ \hline
            split ACP  & 3.1587  \\ \hline
                \end{tabular}
        \end{center}
        \caption{ Computational scaling measured from $N_A = 90$ to $N_A = 140$.}
        \label{tab:1D_time_scaling}
\end{table}

In the end we perform phonon calculations for systems of size from 30 to 140. 
We choose $\delta = 0.01$ for FD approach. Fig.~\ref{fig:1D_phonon:accu} 
shows that the accuracy of phonon spectrum ($L^{\infty}$ error) from FD approach 
remains roughly the same as the system size increases, which is 
empirically around $10^{-4}$. For the split ACP, we find that 
$\epsilon = 10^{-4}$, $N_c = 4$, $N_p = 20$ 
and $\widetilde{N}_{\text{cut}} \approx 1.7N_{\text{cut}}$ is sufficient to achieve 
error around $10^{-4}$. 
Fig.~\ref{fig:1Dphonon:spec} reports the phonon spectrum $\varrho_D$ for 
system of size $N_A = 140$. We remark that Fig~\ref{fig:1Dphonon:spec} plots 
$\varrho_D$ by smearing the Dirac-$\delta$ distribution in~\eqref{eqn:phononspec} 
using a regularized function 
$$\delta_\sigma(x) = \frac{1}{\sqrt{2\pi \sigma^2}} e^{-\frac{x^2}{2\sigma^2}},$$
where the smear parameter $\sigma$ is chosen to be 0.005.

To demonstrate the efficiency of the split ACP formulation, 
Fig.~\ref{fig:1D_time} compares the computational time of different methods. 
We observe that the split ACP can be more advantageous than DFPT 
for systems merely beyond 40 atoms, 
and become more advantageous than FD for systems beyond 60 atoms. 
For the largest system with 140 atoms, split ACP is 3.37 and 1.68
times faster than DFPT and FD, respectively.  

Table~\ref{tab:1D_time_scaling} measures the slope of the computational cost 
with respect to system sizes from $N_A = 90$ to $N_A = 140$. 
In theory, the asymptotic computational cost of DFPT and FD should be $\Or(N_e^4)$, 
and the cost of split ACP should be $\Or(N_e^3)$. 
For all the methods, numerical scalings shown in Table~\ref{tab:1D_time_scaling}
match closely with the theoretical ones.

\subsection{3D aluminum cluster}
In this section, we present the result of phonon calculations of a 3D aluminum cluster. Each unit cell is a $7.65\times7.65\times7.65$ a.u. with 4 Al atoms. 
%which consists of 4 atoms positioned as
% \begin{equation}\label{eqn:alcluster}
% \left[\begin{matrix}
% 0 & 0 & 0 \\
% 3.825 & 3.825 &0 \\
% 0 & 3.825 & 3.825 \\
% 3.825 & 0 & 3.825
% \end{matrix}\right].
% \end{equation}
The computational supercell consists of $2\times 2\times 1$ unit cells and has 16 atoms and $48$ electrons.  We use the spin-restricted formulation and the Perdew-Zunger pseudopotential \cite{PerdewZunger1981}, and the temperature is set to 1000K. $E_\text{cut}$ is set to $10$ Hartree. We set $N_\mathrm{cut} = 33$, $\wt{N}_\mathrm{cut} = 47$, and the number of Chebyshev interpolation $N_c$ to be  6. For the system size tested, we found that using Eq.\eqref{eqn:splitchi0} directly for computing the singular part of the polarizability matrix much more faster than using the pole expansion. So the computation is done using Eq.\eqref{eqn:splitchi0} for the purpose of testing the accuracy of the algorithm. This results in much shorter computational time given the size of the system tested is small.

\begin{figure}[ht]
        \begin{minipage}[h]{\linewidth}
                \centering
                \includegraphics[width=0.7\linewidth]{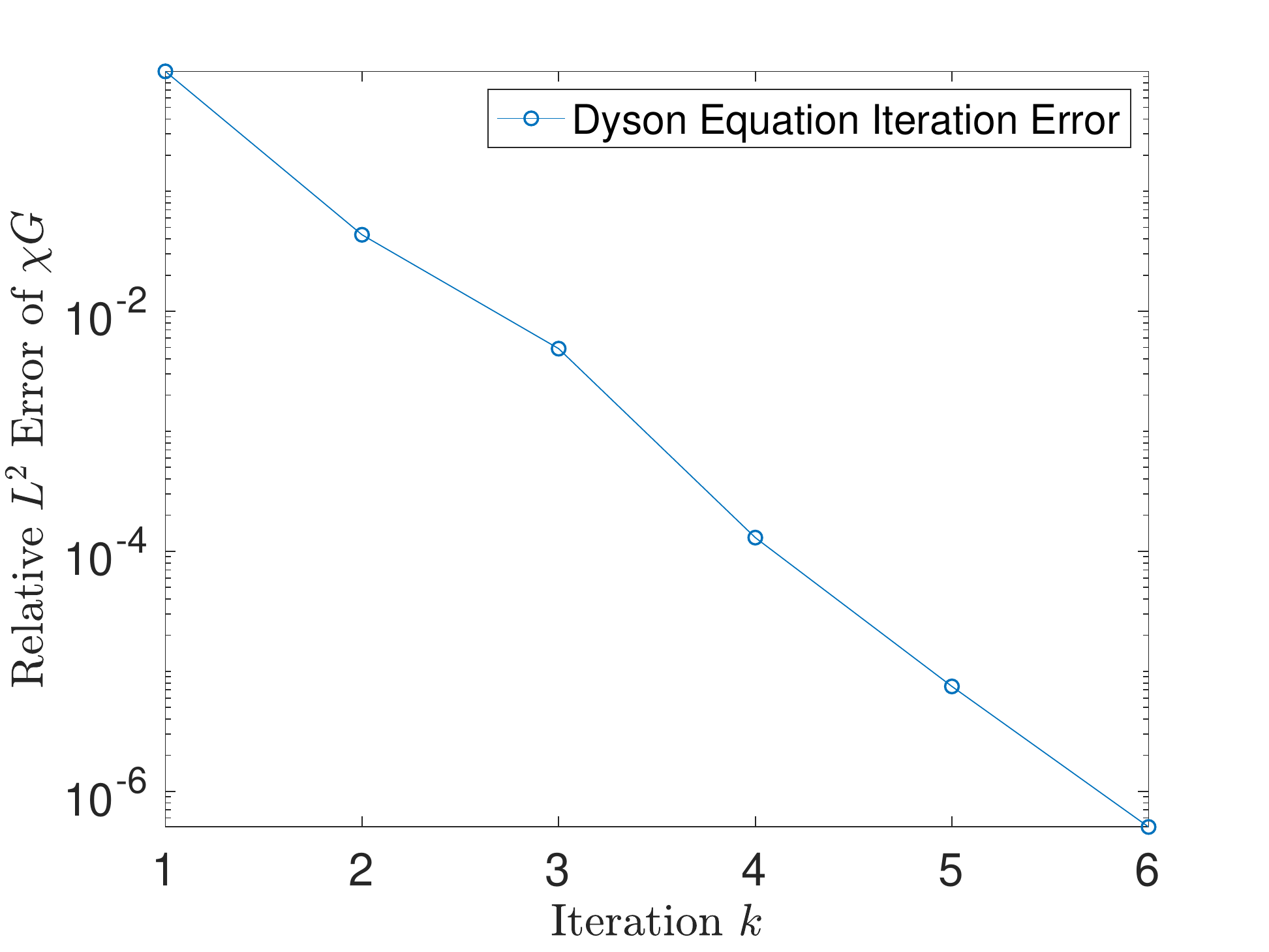}
        \end{minipage}
  \caption{The Dyson Equations iteration error.}
        \label{fig:3D_Dyson}
\end{figure}

Figure~\ref{fig:3D_Dyson} reports the relative error in the iteration of solving the Dyson equation. We remark that for this system, $N_{\mu} = 1584$. In comparison, the total grid points in the discretization is $N_g = 42592$. This means that the numerical rank of the operator $\chi$ far less than the number of grid points. The iteration is converged to $10^{-6}$ relative error for 6 steps. 

\begin{figure}[ht]
        \begin{minipage}[h]{\linewidth}
                \centering
                \includegraphics[width=0.7\linewidth]{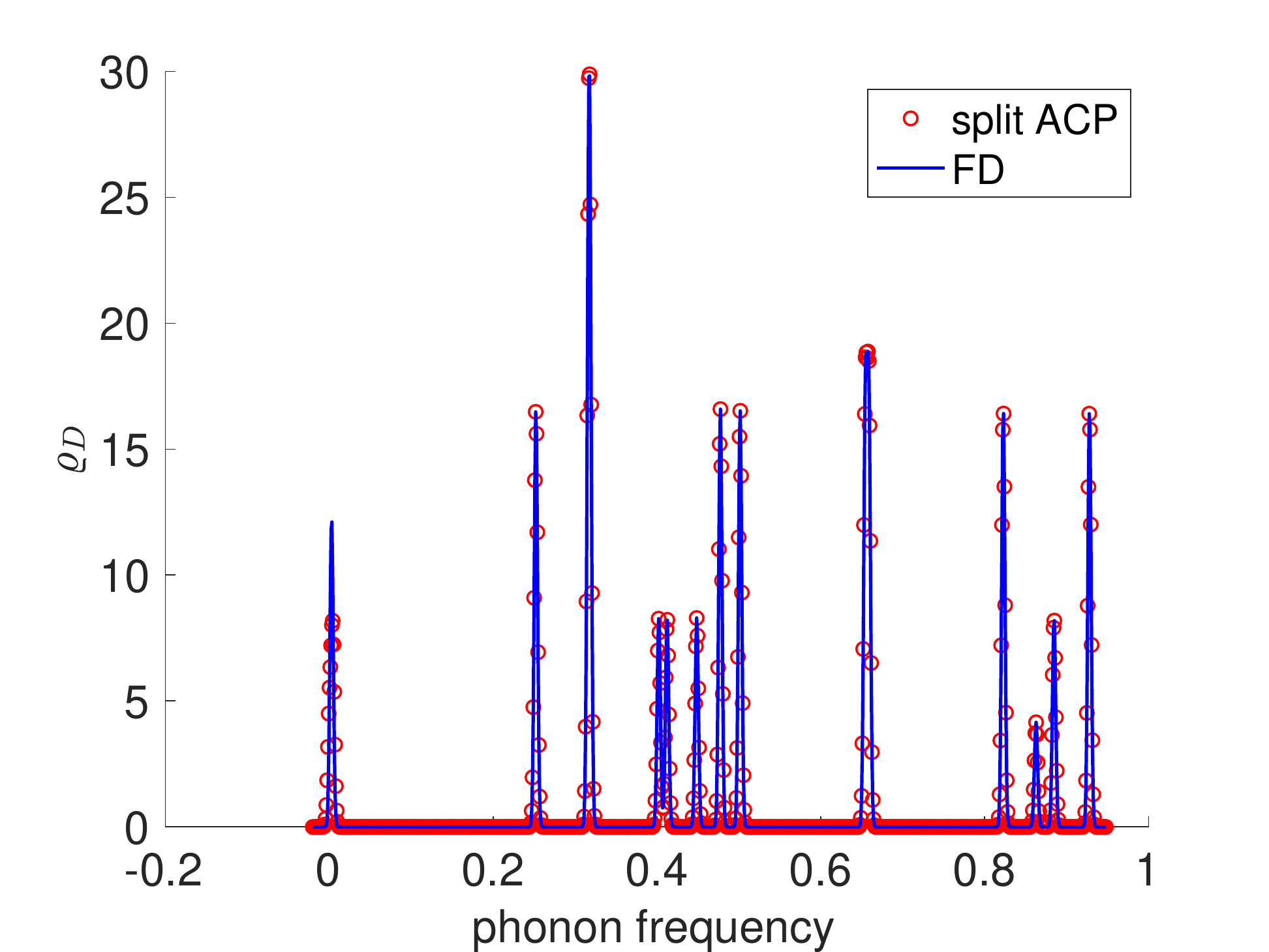}
        \end{minipage}
  \caption{Phonon spectrum of 3D Aluminum Cluster.}
        \label{fig:3D_spec}
\end{figure}

Figure~\ref{fig:3D_spec} reports the phonon spectrum computed from both FD and split ACP. The smearing parameter for plotting the spectrum is chosen as 0.008. 
% Table~\ref{tab:3DPhononError} reports the error on the density of states. 
The $L^{\infty}$ error on the density of states is 5.62E-05.

% \begin{table}[ht]
%         \begin{center}
%                 \begin{tabular}{c|c} \hline
%                         Variable  & $L^\infty$ error ACP vs FD\\ \hline
%             % $\{\omega_k^2\}$  & 3.58E-05 \\ \hline
%             % $\{\omega_k\approx 0\}$    & 3.32E-03 \\ \hline
%             $\{\omega_k>0\}$  & 5.62E-05  \\ \hline
%                 \end{tabular}
%         \end{center}
%         \caption{ Error on density of states and eigenvalues of dynamical matrix.  \LL{Remove the $\omega_k\approx 0$ part. Also why showing $\omega_k^2$ and $\omega_k$ together?} }
%         \label{tab:3DPhononError}
% \end{table}

We remark that the purpose of the test above is to illustrate that the split ACP formulation can indeed be used to accurately obtain the phonon spectrum for 3D metallic systems, with fractionally occupied states and nonlocal pseudopotentials. However, due to the small system size, the computational time of the split representation of ACP is in fact much longer than that of FD. Also we remark that there is difficulty in the DFPT approach in 3D. The Sternheimer equations are ill-conditioned and the MINRES iteration fail to converge. This result also emphasizes the necessity of introducing the effective gap in the split ACP. 

Since KSSOLV is only designed to solve Kohn-Sham equations for systems with relatively small sizes, our implementation cannot reveal the efficiency of the split ACP approach yet for 3D systems, and this will be our future work.
\section{Conclusion}\label{sec:conclusion}

We have introduced the split representation of a recently developed method called the adaptively compressed polarizability operator. The split ACP formulation incorporates nonlocal pseudopotentials and finite temperature effects successfully, hence generalizes the ACP formulation to solve for phonons in metallic systems as well. Our numerical results for model problems indicate that the computational advantage of the split ACP fomulation can be clealy observed compared to DFPT and finited difference, even for systems of relatively small sizes. The numerical example for 3D Aluminum cluster shows that accuracy of the split ACP formulation in the application for computing the phonon spectrum for real materials.

The new split representation of ACP provides a systematic and complete solution to treating systems at finite temperature. We have used phonon calculation as an example to demonstrate the effectiveness as well as accuracy of the split representation of adaptively compressed polarizability operator. The same strategy can be applied to applications of DFPT other than phonon calculations, when the polarizability operator $\chi$ needs to be applied to a large number of vectors. Moreover, Meanwhile, all numerical tests are on single-threaded. Parallelized
implementation would help fully test whether split representation of ACP formulation can
achieve the goal of reducing complexity to asymptotically $\Or(N_e^3)$. We will present the parallel implementation in the future.

\section*{Acknowledgments} 
This work was partially supported by the National Science
Foundation under Grant No. DMS-1652330 (D. A. and L. L.), the U.S. Department of Energy under Contract No.
DE-SC0017867 (L. L. and Z. X.), and the  U.S. Department of Energy under the Center for Applied
Mathematics for Energy Research Applications (CAMERA) program (L. L.). We thank Berkeley Research Computing for the computational resources.

\section*{Appendix A}

Using the Cauchy contour integral formulation, the density matrix at finite temperature can be represented as
\begin{equation}
        P_0 = \frac{1}{ 2\pi i }  \oint_{\mc{C}} f(z) (z - H)^{-1} \ud z. 
\end{equation}
When the Hamiltonian is perturbed to $ H_{\varepsilon} = H_{0} + \varepsilon \mf{g}$, and when $\varepsilon$ is small enough, the perturbed density matrix $P_\varepsilon$ can still be computed as
\begin{equation}
        P_\varepsilon = \frac{1}{ 2\pi i }  \oint_{\mc{C}} f(z) (z - H_{\varepsilon})^{-1} \ud z. 
\end{equation}
Then we have
\begin{equation}
        \begin{split}
        P_{\varepsilon} - P_0 &= \frac{1}{2\pi i} \oint_{\mc{C}}   f(z) \left[ (z - H_{\varepsilon})^{-1}  - (z - H)^{-1}  \right] \ud z \\
        &= \frac{1}{2\pi i} \oint_{\mc{C}}   f(z) \left[ (z - H_{\varepsilon})^{-1} \varepsilon \mf{g} (z - H)^{-1}  \right] \ud z \\
        &= \frac{1}{2\pi i} \oint_{\mc{C}}   f(z) \left[ (z - H)^{-1} \varepsilon \mf{g} (z - H)^{-1}  \right] \ud z + \Or(\varepsilon^2).
        \end{split}
\end{equation}

Hence by the definition of $\mf{X}_{0}$, we have
\begin{equation}
        \mf{X}_{0} \mf{g} = \frac{1}{2\pi i} \oint_{\mc{C}}   f(z) \left[ (z - H)^{-1} \mf{g} (z - H)^{-1}  \right] \ud z.
\end{equation}
Using the spectral decomposition of $H$, and use the contour integral formulation
\begin{equation}
  \begin{split}
  \mf{X}_{0} \mf{g} &= \frac{1}{2\pi i} \oint_{\mc{C}}  \sum_{j,k = 1}^{\infty} f(z) \left[ \frac{\psi_j \psi_j^* \mf{g} \psi_k \psi_k^*}{(z-\varepsilon_j)(z-\varepsilon_k)}  \right] \ud z \\
  &= \frac{1}{2\pi i} \sum_{j,k = 1}^{\infty} \oint_{\mc{C}}  \ud z\frac{f(z) }{(z-\varepsilon_j)(z-\varepsilon_k)} \left[   \psi_j \psi_j^* \mf{g} \psi_k \psi_k^*\right] \\
  &= \sum_{j\ne k}^{\infty} \frac{f_j - f_k }{\varepsilon_j-\varepsilon_k} \left[   \psi_j \psi_j^* \mf{g} \psi_k \psi_k^*\right] + \sum_{j}^{\infty} f'_j \left[   \psi_j \psi_j^* \mf{g} \psi_j \psi_j^*\right] \\
  &= \sum_{j, k}^{\infty} \frac{f_j - f_k }{\varepsilon_j-\varepsilon_k} \left[   \psi_j \psi_j^* \mf{g} \psi_k \psi_k^*\right],
  \end{split}
\end{equation}
where the $\frac{f_j - f_k}{\varepsilon_j - \varepsilon_k}$ is interpreted as the derivative when $j=k$.

For the purpose of computing singular part with contour representation, we have

\begin{equation}
  \begin{split}
  \mf{X}_{0}^{(s)} \mf{g} =&\sum_{i=1}^{N_{\mathrm{cut}}}  \sum_{a=N_{\mathrm{cut}} +1}^{\wt{N}_{\mathrm{cut}}}
  \frac{f_{a}-f_{i}}{\varepsilon_{a}-\varepsilon_{i}}
  \psi_{a}(\psi_{a}^{*} \mf{g} \psi_{i})\psi_{i}^{*} + \mathrm{h.c.}   \\
  & +  \sum_{i=1}^{N_{\mathrm{cut}}}  \sum_{a=1}^{N_{\mathrm{cut}}}
  \frac{f_{a}-f_{i}}{\varepsilon_{a}-\varepsilon_{i}}
  \psi_{a}(\psi_{a}^{*} \mf{g} \psi_{i})\psi_{i}^{*}  \\
  =& \frac{1}{2\pi \I} \oint_{\mc{C}} \ud z  \sum_{i=1}^{N_{\mathrm{cut}}}  \sum_{a=N_{\mathrm{cut}}+1}^{\wt{N}_{\mathrm{cut}}} \frac{f(z)}{(z-\varepsilon_a)(z-\varepsilon_i)} \left[  \psi_a \psi_a^* \mf{g} \psi_i \psi_i^*\right] + \mathrm{h.c.}\\
  &+  \frac{1}{2\pi \I} \oint_{\mc{C}} \ud z  \sum_{i=1}^{N_{\mathrm{cut}}}  \sum_{a=1}^{N_{\mathrm{cut}}} \frac{f(z)}{(z-\varepsilon_a)(z-\varepsilon_i)} \left[  \psi_a \psi_a^* \mf{g} \psi_i \psi_i^*\right] \\
  =& \frac{1}{2\pi \I} \oint_{\mc{C}} f(z)
        (z-H_{c,2})^{-1}\mf{g}(z-H_{c,1})^{-1} \ud z  + \mathrm{h.c.} \\
         & + \frac{1}{2\pi \I} \oint_{\mc{C}} f(z)
        (z-H_{c,1})^{-1}\mf{g}(z-H_{c,1})^{-1} \ud z,
  \end{split}
\end{equation}
where $H_{c,1}=\sum_{i=1}^{N_{\mathrm{cut}}} \psi_{i}
\varepsilon_{i}\psi_{i}^{*}, H_{c,2}=\sum_{i=N_{\mathrm{cut}}+1}^{\wt{N}_{\mathrm{cut}}} \psi_{i}
\varepsilon_{i}\psi_{i}^{*}$ are the Hamiltonian operators projected to
the subspace spanned by the first $N_{\mathrm{cut}}$ states, and to
the subspace spanned by the following $(\wt{N}_{\mathrm{cut}} - N_{\mathrm{cut}})$ states, respectively.

%%%%%%%%%%%%%%%%%%%%%%%%%%%%%%%%%%%%%%%%%%%%%%%%%%%%%%%%%%%%%%%%%%%%%%%%%%%%%%%%

%%%%%%%%%%%%%%%%%%%%%%%%%%%%%%%%%%%%%%%%%%%%%%%%%%%%%%%%%%%%%%%%%%%%%%%%%%%%%%%%
\bibliographystyle{siam}
\bibliography{reference}

\end{document}